\documentclass[traditabstract]{aa}
\usepackage{graphicx}
\usepackage{amsmath,amssymb}
\usepackage{natbib}
\bibpunct{(}{)}{;}{a}{}{,}

\newcommand{\Teff}{T_\mathrm{eff}}

\begin{document}
   \title{Towards a new generation of multi-dimensional stellar evolution models: development of an implicit hydrodynamic code}
\author{M. Viallet\inst{1,2}, I. Baraffe\inst{2,1}, R. Walder\inst{1}}

   \institute{ \'Ecole Normale Sup\'erieure, Lyon, CRAL (UMR CNRS 5574), Universit\'e de Lyon 1, France\\
%	\institute{Universit\'e  de Lyon, Lyon, F-69003, France; Ecole Normale Sup\'erieure de Lyon, 46 all\'ee dÕItalie, Lyon, F-69007, France; CNRS, UMR 5574, Centre de Recherche Astrophysique de Lyon; Universit\'e Lyon 1, Villeurbanne, F-69622 France\\
             \email{maxime.viallet@ens-lyon.fr}
             \and
             Physics and Astronomy, University of Exeter, Stocker Road, Exeter, UK EX4 4QL
             }

   \date{Received; accepted}

  \abstract{This paper describes the first steps in the development of a new multi-dimensional time implicit code devoted to the study of hydrodynamical processes in stellar interiors. The code solves the hydrodynamical equations in spherical geometry and is based on the finite volume method. Radiation transport is taken into account within the diffusion approximation. Realistic equation of state and opacities are implemented, allowing study of a wide range of the problems characteristic of stellar interiors. We describe the numerical method and various standard tests performed to validate the method in detail. We present preliminary results devoted to describing stellar convection. We first performed a local simulation of convection in the surface layers of a A-type star model. This simulation tested the ability of the code to address stellar conditions and to validate our results, since they can be compared to similar previous simulations based on explicit codes. We then present a global simulation of turbulent convective motions in a cold giant envelope, covering 80\% in radius of the stellar structure. Although our implicit scheme is unconditionally stable, we show that in practice there is a limitation on the time step that prevents the flow moving over several cells during a time step. Nevertheless, in the cold giant model we reach a hydro CFL number of 100. We also show that we are able to address flows with a wide range of Mach numbers ($10^{-3} \lesssim \mathrm{M}_s \lesssim 0.5$), which is impossible with an anelastic approach. Our first developments are meant to demonstrate that applying an implicit scheme to a stellar evolution context is perfectly thinkable and to provide useful guidelines for optimising the development of an implicit multi-dimensional hydrodynamical code.}

     
  \titlerunning{Implicit calculations of stellar interiors}
  \authorrunning{M. Viallet et al.}
   
  \maketitle

%
%________________________________________________________________

\section{Introduction}
\label{introduction} 

With the advent of massively parallel computers and the development of advanced numerical techniques like adaptive mesh refinement and massive parallelisation, astrophysical fields like cosmology, star formation, or stellar/galactic environment studies have recently made remarkable steps forward.  
Despite this revolution in computational astrophysics, our physical understanding of stellar interiors and evolution still largely relies on one-dimensional calculations. Implicit 1D stellar evolution codes were widely developed during the past century, gaining more and more in sophistication through improvements in the input physics and implementation of an increasing number of complex physical processes such as time-dependent turbulent convection, rotation, or MHD processes. Their description, however, still mostly relies on simplified, phenomenological approaches, with a predictive power hampered by the use of several free parameters. 
These approaches have now reached their limits for understanding stellar structure and evolution. The development of multi-dimensional hydrodynamical simulations becomes crucial for progress in the field of stellar physics and for the enormous observational efforts aimed at producing data of unprecedented quality, as expected from  the asteroseismological space projects COROT and Kepler. 

Several efforts have been devoted to the development of 2D and 3D tools for studying some aspects and specific problems of stellar structure and evolution. Impressive progress and developments have been achieved in multi-dimensional hydrodynamics and MHD models of stellar atmospheres and photospheres \citep[see e.g.][]{1996A&A...313..497F,1998ApJ...499..914S,2000A&A...359..669A,2006A&A...446..635B,2009AIPC.1171..242N}. In stellar interiors, many studies of convection, rotation, and magnetic fields are performed with anelastic hydrodynamic solvers, which filter out  sound waves and have the priceless advantage of not being restricted to the Courant-Friedrich-Lewy (CFL hereafter) time step limit, a severe limitation for explicit compressible hydrodynamic solvers \citep[][and references therein]{2002ApJ...570..865B,2010arXiv1005.5406G}.
%(Brun \& Toomre 2002; Rogers et al. 2010; see Guzik 2010 and reference therein). 
Anelastic approaches, however, are restricted to the study of flows characterised by very low Mach numbers and small thermodynamic fluctuations from the background \citep[see discussions in][]{2006ApJ...637..922A, 2007ApJ...665..690M}. 

Other groups use 2D/3D explicit hydrodynamical simulations to follow the properties on short snapshots of turbulent convection, mixing processes, or nuclear burning in stellar interiors such as cores and burning shells of massive stars or during the core He flash in low-mass stars \citep{2006ApJ...639..405D, 2007ApJ...667..448M, 2008ApJ...677..581E, 2009A&A...501..659M,2009ApJ...690.1715A}. However, owing to the intrinsic time-step limitation of explicit codes, simulating those processes on time scales that are significantly larger than the dynamical one and relevant to stellar evolution would be prohibitively expensive. To overcome this problem, increasing efforts are now devoted to developing algorithms for low Mach-number, compressible stellar flows \citep{2009MNRAS.400..903H,2010ApJS..188..358N}. However, many stellar regions and evolutionary phases are characterised by low-to-moderate Mach-number flows, which would ideally be described by fully time-implicit solvers. 

We only know two multi-dimensional implicit codes applied to stellar physics problems, namely the 2D stellar evolution code ROTORC developed by \citet{1990ApJ...357..175D} and the 2D code VULCAN of \citet{1993ApJ...412..634L}. The work of Deupree is pioneering in the field of multi-dimensional implicit stellar evolution, with ROTORC applied to the study of stellar convection and rotation in stars of different masses. However,  the numerical method involved in this code cannot compete with current, advanced numerical algorithms, because it is based on a non-conservative finite-difference scheme with the implicit solver as an extension to the 2D of the Henyey method. This method is widely used in 1D stellar evolution calculations but cumbersome (if not impossible) to parallelise, thus it significantly restricts the spatial resolution for multi-dimensional calculations. VULCAN is based on a two-step method consisting of an implicit Lagrangian step followed by an explicit remapping step on an arbitrary grid. The code has been essentially applied to studying core-collapse supernovae \citep[see e.g.][]{2004ApJ...609..277L,2007PhR...442...23B} and novae explosions \citep[see e.g.][]{1995ApJ...445L.149G,1995ApJ...452...62L,1997ApJ...475..754G,1999ApJ...527L..97L}. Although VULCAN offers the possibility of solving advection implicitly, to our knowledge in these works the authors performed multi-dimensional computations by solving advection explicitly and radiation implicitly. Thus, we cannot gauge the capacity of this code to describe long-term stellar evolution problems, in general, and convective flows in various stellar interiors, in particular. \citet{2005A&A...430..893H,2005CoPhC.168....1H} develops a multi-dimensional implicit MHD solver, but the applications are oriented more towards high-energy processes (e.g. accretion on a black hole). Finally, we mention the work of \cite{1998A&A...332.1159T}  and \cite{so66811}, which describe the implementation and tests of various implicit methods for multi-dimensional computations of steady state and time-dependent problem in (magneto)hydrodynamic.

In this context, we started to develop a multi-dimensional time implicit code, guided by the motivation of improving the description of stellar interiors during different phases of evolution. The holy grail is to develop a tool that allows the 3D description of a complete star during  thermal, nuclear, or accretion time scales that would be relevant to various stellar evolution phases.  Our primary interests focus on the description of time-dependent convection in the envelope of pulsating stars like Cepheids, a long-standing problem in asteroseismology, or during the very early phases of stellar evolution (early pre-main sequence) following the hydrodynamical phase of a proto-star formation. Improving our description of these very early stages of evolution is crucial to a more accurate analysis of current observations and, in particular, to derivation of a key property for understanding star formation, namely the initial mass function (see e.g discussion in \citealt{2002A&A...382..563B}). For the above-mentioned problems, neither an anelastic nor a low-Mach approach is appropriate,  since the convective flows are characterised by very low ($M \ll 0.01$) to moderate ($0.1 \lesssim M \lesssim 1$) Mach numbers from the deep interior to  the stellar surface. Also, a wide range of other astrophysicaly interesting problems can be addressed with such a tool, providing a wealth of applications on a long-term future and a revival of the field of stellar physics. 

Our goal is exceedingly challenging, with a long way to go before achievement. This first paper describes our method and the first tests performed to describe convection in stellar interiors. Because implicit schemes are much more expensive in terms of CPU than explicit ones, part of the astrophysical, computational community casts doubts on the advantage and applicability of the former to any stellar physics problem. Our first developments are meant to demonstrate that the use of an implicit scheme in a stellar evolution context is perfectly thinkable. For this purpose, we present two examples of stellar convection for which we derive the exact CFL numbers and discuss the advantages and limits of an implicit approach. Even though the implicit method used throughout the paper has no stability limit on the time step (see Sect. \ref{implicitstrategy}), the time step cannot take arbitrarily high values for both technical and accuracy reasons. We show in this paper that both reasons are actually related and that the time step cannot grow much larger than the shortest time scale of the flow for crossing a grid cell. Technically this implies that the solution change is moderate between two time steps, insuring a safe convergence in the non-linear solver. In addition, this is also a good accuracy criterion, since advection over more than one grid cells is subject to strong numerical damping. Our first numerical experiments are meant to provide useful guidelines for developing an implicit multi-dimensional hydrodynamical code.

The paper is organised as follows. We describe in Sect. \ref{section:Equations} our physical model and in Sect. \ref{section:numerics} our numerical method. We present a test case in Sect. \ref{section:tests} and results for stellar models in Sect. \ref{section:stars}. In Sect. \ref{conclusion}, we conclude the paper and discuss future numerical advancements that are planned to optimise the code.
 
 %__________________________________________________________________

\section{Equations}
\label{section:Equations}

We solve the equations describing the evolution of density, momentum, and internal energy, taking into account external gravity and radiative diffusion:

\begin{eqnarray}
\frac{\partial}{\partial t} \rho &=& - \vec \nabla . (\rho \vec u)\\
\frac{\partial}{\partial t} \rho e &=& -\vec \nabla . (\rho e \vec u) - P\vec \nabla . \vec u - \vec \nabla . \vec F_r + \frac{1}{2\rho \nu}\tau_{ij} \tau^{ij} \\
\frac{\partial}{\partial t} \rho \vec u &=& - \vec \nabla . (\rho \overline{\overline{uu}}-\overline{\overline{\tau}})-\vec \nabla P + \rho \vec g,
\end{eqnarray}

\noindent where $\rho$ is the density, $e$ the specific internal energy, $\vec u$ the velocity, $P$ the gas pressure, $F_r$ the radiative flux (see below), $\vec g$ the gravitational acceleration, and $\overline{\overline{\tau}}$ the viscous stress tensor, whose components are given by

\begin{equation}
\tau_{ij} = 2\rho \nu \Big( e_{ij} - \frac{1}{3} \delta_{ij} \vec \nabla . \vec u \Big),
\end{equation}

\noindent where $e_{ij} $ is the strain rate tensor and $\nu$ the viscosity coefficient. The expected value of the molecular viscosity in stellar interiors implies large Reynolds numbers characterising the flow (up to $10^{12}$). It is therefore impossible to model all scales of the flow, from the stellar scale down to the dissipation scale, on current generation of  computers. 
As a result, any numerical simulations of stellar interiors should be interpreted in the large eddy simulation (LES) paradigm, in which only large-scale motions are resolved on the grid. In the standard LES, the effect of subgrid-scale motions (i.e. turbulence) is taken into account by introducing a so-called subgrid scale (SGS) model. A classical example is the effective viscosity coefficient proposed in \cite{Smagorinsky_1963}. On the other hand, the MILES approach (for Monotone Integrated Large Eddy simulation, see e.g. \citealt{1873-7005-10-4-6-A01}) relies solely on the numerical viscosity of the scheme, which results from discretisation, to mimic the subgrid scale dissipation. In our case, a standard LES approach should rely on an SGS model relevant for compressible flow in a stratified medium, which to our knowledge has not been designed yet. It is far from certain that the use of standard SGS models, mostly relevant for incompressible and homogenous flow, will improve our results. Therefore, we choose to follow the MILES strategy and all the computations of stellar convection presented in this paper are based on the advection-diffusion equations without explicit viscosity.

%Another reason to introduce a larger value of $\nu$ is numerical robustness, an artificial viscosity (see e.g. REF) can be introduced to handle shocks that can develop in some situations. In the rest of the paper, we will however set the viscosity to zero (MORE JUSTIFICATIONS HERE).

We solve the internal energy density equation. Some of the popular numerical codes that have been used to study stellar convection solve the entropy equation (e.g. ASH code, Pencil code), which is equivalent  to the internal energy equation. A  better approach to energy conservation would be to solve the total energy equation, as for instance in the COBOLD code (see Freytag and collaborators) and in the PROMPI code (see \citealt{2007ApJ...667..448M}). In the future we plan to implement a total energy equation solver, which might be more appropriate for problems like stellar pulsations.

For stellar calculations, we treat radiation transport in the diffusion limit approximation. This approximation is suitable for an optically thick region but becomes inaccurate in the photosphere and optically thin regions. In this framework, the radiative flux $F_r$ writes as

\begin{equation}
\vec F_r = - k_\mathrm{rad} \vec \nabla T,
\end{equation}

\noindent where the photon conductivity $k_\mathrm{rad}$ is given by

\begin{equation}
k_\mathrm{rad} = \frac{4acT^3}{3\kappa \rho},
\end{equation}

\noindent with $\kappa$ the Rosseland opacity of the gas, which is interpolated from the OPAL tables \citep{1996ApJ...464..943I}.

The equations are closed with the equation of state:

\begin{equation}
P=P(\rho,e)\mathrm{,\ \ \ } T=T(\rho,e).
\end{equation}

\noindent Our equation of state is tabulated and includes the partial ionisation of several chemical elements (determined by the Saha-Bolztmann equations) from hydrogen to silicium and takes Coulombian pressure into account. The chemical abundances are set to the solar values. For details see \cite{1996A&A...312..845A}. We consider here stellar envelopes and therefore we do not take nuclear reactions into account, which concern the central region of the star.

For the geometry, we adopt spherical coordinates. We assume azimuthal symmetry so the only independent coordinates are $r$ (the radius) and $\theta$ (the colatitude). In general, the coordinates $(r,\theta)$ span a domain $[r_\mathrm{in},r_\mathrm{out}]\times[\theta_1,\theta_2]$. We use periodic conditions in the tangential direction. For boundary conditions at the top and bottom of the domain we impose

\begin{enumerate}
\item non-penetrative conditions: $u_r = 0 $ at $r=r_\mathrm{in}$,$r_\mathrm{out}$;
\item stress-free conditions:  $\frac{\partial}{\partial r} \Big ( \frac{u_\theta}{r}\Big )$ at $r=r_\mathrm{in}$,$r_\mathrm{out}$;
\item constant energy flux $F_\star$ at the bottom of the domain $r=r_\mathrm{in} $;
\item an energy flux at the top of the domain given by
\begin{equation}
	F_\mathrm{out} = \sigma T_\mathrm{out}^4,
\end{equation}
\end{enumerate}

\noindent where $T_\mathrm{out}$ is the temperature of the cells in the top layer of the domain. %Note that this prescription for the outgoing energy flux ensures that in steady state the top layer of cells has a temperature equal to the star effective temperature.
Finally, we would like to stress that we adopted 2D geometry to simplify the development of the code, but extension to 3D is planned in the future.

\section{Numerical method}
\label{section:numerics}

We present below our numerical method, describing separately  the spatial (Sect. \ref {spatial}) and temporal (Sect. \ref{temporal}) discretisations.

\subsection{Spatial discretisation}
\label{spatial}

\subsubsection{One-dimensional advection scheme}

The equations are discretised on a staggered grid, using a finite volume approach. 
Cell interfaces location are denoted by $x_i,i=1\dots N_x$ and  cell centres by $x_{i+1/2},i=1\dots N_x-1$. In the staggered grid approach, scalar quantities (e.g. density, internal energy, pressure, temperature, etc) are located at cell centres, while velocity components are located at cell boundaries.

In the finite volume approach, the physical equations are integrated over a control volume to yield the evolution law for the physical quantities (mass, internal energy, momentum) in each cell. Fluxes are computed at cell interfaces, yielding a conservative scheme with respect to advection. For scalar quantities (density and internal energy) at $x_{i+1/2}$, the control volume is the cell $[x_i,x_{i+1}]$ with volume $V_{i+1/2}$ and surfaces $S_i$, $S_{i+1}$. For the velocity component at $x_i$, the control volume is the cell $[x_{i-1/2},x_{i+1/2}]$, with volume $\hat{V}_{i}$ and surfaces $\hat{S}_{i-1/2}$, $\hat{S}_{i+1/2}$. 
 
We use the Van Leer method to compute the fluxes (see \citealt{1977JCoPh..23..276V,1992ApJS...80..753S}). To compute the flux at the control volume interfaces, it is necessary to perform interpolation since not all quantities are available at the considered interface. We define ``barred" quantities (i.e $\bar{\rho}$, $\bar{u}$) as quantities that are interpolated at the interface by taking a simple average between the values available on each side of the interface, for instance $\bar{u}_{i+1/2} = (u_i + u_{i+1})/2$. For a stable and TVD advection, it is necessary to introduce some limiting process in the interpolation scheme of advected quantities. Here we proceed with an upwind limited interpolation similar to the MUSCL method. In each cell a linear reconstruction of the solution is performed. For instance, we describe below the procedure for cell $i+1/2$ and cell centred variable $q$. We first compute the limited slope $\Delta_{i+1/2} q$ of the linear reconstruction in the cell by using the values of the neighbouring cells with a limiting process:

\begin{eqnarray}
\Delta_{i+1} q &=& q_{i+3/2} - q_{i+1/2}\\
\Delta_{i} q &=& q_{i+1/2} - q_{i-1/2}\\
\Delta_{i+1/2} q &=& \phi(\frac{\Delta_{i+1} q}{\Delta_{i} q}) \Delta_{i} q,
\end{eqnarray}

\noindent where $\phi$ is the slope limiter. In the following we use the Van Leer limiter:

\begin{eqnarray}
\phi(r) &=& \frac{2r}{1+r} \mathrm{\ \ \ if\ r>0}\\
           &=& 0 \mathrm{, \ otherwise.}
\end{eqnarray}

With such a reconstruction scheme, we reconstruct at each interface right and left values $q^R$ and $q^L$:

\begin{eqnarray}
q^R_i &=& q_{i+1/2} - \Delta_{i+1/2} q\\
q^L_i &=&q_{i-1/2} + \Delta_{i-1/2} q.
\end{eqnarray}

\noindent Depending on the sign of the velocity, one of these quantities is used to compute the flux at the interface. For instance, at interface $x_i$ one has

\begin{eqnarray}
	\tilde{q}_i & = & q^L_i \mathrm{\ if\ u_i \ge 0}\\
	      	     & = & q^R_i   \mathrm{\ if\ u_i < 0}.
\end{eqnarray}

\noindent This defines the ``tilded" quantities in the equations shown below. For the interfaces located at $i+1/2$, the same reconstruction is performed and the interface value is chosen depending on the sign of $\bar{u}_{i+1/2}$.

The pressure gradient at the interface is computed with a second-order, centred difference. The overall scheme is expected to be second order in space in the smooth regions of the flow. We present in appendix \ref{appendix:tests} an order study for the linear advection problem.

After integration over the corresponding control volume and discretisation, the 1D Euler equations become

\begin{align}
\label{ODEs}
% Density
\frac{d}{d t} \rho_{i+1/2} V_{i+1/2} = & - \big( S_{i+1} \tilde{\rho}_{i+1} u_{i+1} - S_{i} \tilde{\rho}_{i} u_{i} \big ) \\
% Internal energy
 \frac{d}{d t} \rho_{i+1/2} e_{i+1/2}V_{i+1/2} = &- \big( S_{i+1} \widetilde{\rho e}_{i+1} u_{i+1} - S_{i} \widetilde{\rho e}_{i} u_{i} \big ) \nonumber \\
% &&+ \frac{ac}{3} \Big ( \frac{S_{i+1} }{\overline{\rho \kappa}_{i+1}} \frac{T_{i+3/2}^4-T_{i+1/2}^4}{\Delta x} - \frac{S_{i} }{\overline{\rho \kappa}_{i}} \frac{T_{i+1/2}^4 -T_{i-1/2}^4}{\Delta x} \Big )\\ \nonumber
& - P_{i+1/2} (S_{i+1} u_{i+1} - S_{i} u_i)\\
% Momentum
\frac{d}{d t} \bar{\rho}_{i} u_{i} \hat{V}_{i} = & - \big( \hat{S}_{i+1/2} \widetilde{\rho u}_{i+1/2} \bar{u}_{i+1/2} \nonumber \\
&- \hat{S}_{i-1/2}\widetilde{\rho u}_{i-1/2} \bar{u}_{i-1/2} \big ) \nonumber \\
& - \frac{P_{i+1/2} - P_{i-1/2}}{\Delta x}\hat{V}_{i}.
 \end{align}
 
\subsubsection{Gravity}

Since we are solving the internal energy equation, gravity only enters the momentum equation. We assume that gravity remains \emph{constant} and \emph{spherically symmetric} so that the gravitational acceleration at radius $r_i$ is given by

\begin{equation}
g_i = - \frac{G M_i}{r_i^2} ,
\end{equation}

\noindent where $G$ is the gravitational constant and $M_i$ the mass contained inside the radius $r_i$. Our stellar models do not extend to the centre of the star so $M_i$ is computed assuming $M_1=M_\mathrm{core}$, where  $M_\mathrm{core}$ is determined from the original stellar model. 

Under the assumption of spherical symmetry, the gravitational force enters the radial momentum equation only through the additional source term:

\begin{equation}
F^\mathrm{grav}_{i} = \bar{\rho}_{i} g_{i} \hat{V}_{i}.
\end{equation}

\noindent The assumption of spherical symmetry could become invalid if the star developed significant non-radial oscillations. The convective flows, which are not symmetric, do not alter the mass distribution symmetry significantly. Implementation of a full solver for the Poisson equation is planned for the future. 

\subsubsection{Radiative diffusion}

The radiative flux through the cell interface is computed with a second-order central difference

\begin{equation}
F^r_i = \frac{ac}{3} \frac{1}{\overline{\rho \kappa}_{i}} \frac{T_{i+1/2}^4 -T_{i-1/2}^4}{\Delta x},
\end{equation}

\noindent where

\begin{equation}
\frac{1}{\overline{\rho \kappa}_{i}} = \frac{1}{2} \Big ( \frac{1}{\rho_{i+1/2} \kappa_{i+1/2}}  + \frac{1}{\rho_{i-1/2} \kappa_{i-1/2}} \Big ).
\end{equation}

\subsubsection{Extension to 2D and spherical geometry}

\begin{figure}[t] %  figure placement: here, top, bottom, or page
   \centering
   \includegraphics[width=0.6\linewidth]{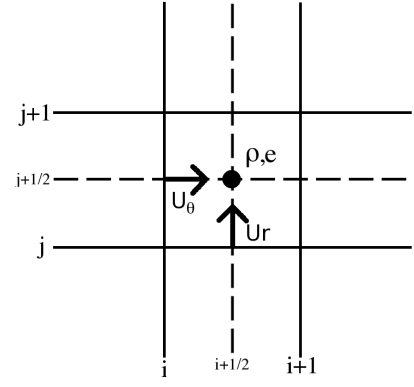} 
   \caption{Staggered mesh in 2D}
   \label{fig:staggered_mesh}
\end{figure}

Extension to 2D is done in the simplest possible way. Fluxes are computed along both directions following the 1D algorithm sketched above and added together in the update step (see below). 
Figure \ref{fig:staggered_mesh} shows the location of all quantities on our staggered 2D mesh. We avoid dimensional splitting (e.g. Strang splitting) that would increase the cost of the solver since several (implicit) substeps would have to be computed to advance the solution over one time step. It is clear in this case that the accuracy of the method as deduced from 1D analysis does not extend to multi-dimensional computation. One solution to improve the accuracy of the code would be to implement a genuinely multi-dimensional method (see e.g. \citealt{Colella:1990:MUM:78582.78634,LeVeque1997327}). Another possibility would be to increase the spatial order of the 1D reconstruction scheme to obtain a better accuracy in multi-dimensional computations. This is left for future developments.

We describe below how spherical geometry is taken into account in our spatial discretisation:
in our finite volume approach, the cell $(i+1/2,j+1/2)$ spans the parameter space domain $[r_i,r_{i+1}]\times [\theta_j, \theta_{j+1}]$. The geometrical factors in the 2D version of cell-centred equations reads as

\begin{eqnarray}
V_{i+1/2,j+1/2} &=& \frac{2\pi}{3}(r_{i+1}^3 - r_i^3)(\cos \theta_j - \cos \theta_{j+1})\\
S^r_{i,j+1/2} &=& 2 \pi r_i^2(\cos \theta_j - \cos \theta_{j+1})\\
S^\theta_{i+1/2,j} &=& \pi \sin \theta_j (r_{i+1}^2-r_{i}^2).
\end{eqnarray}

These geometrical factors account for the geometrical effect associated with spherical coordinates.
There are similar expression for the radial momentum and the tangential momentum equations (see appendix \ref{appendix:formulae}); however, due to the vector nature of the velocity, there are additional geometric source terms that appear in the momentum equations. In the radial momentum equation, the geometric source term reads as

\begin{equation}
S^\mathrm{geom}_{i,j+1/2} = \hat{V}_{i,j+1/2} \frac{ \bar{\rho}_{i,j+1/2} (\bar{u}^\theta_{i,j+1/2})^2 }{r_i},
\end{equation}

\noindent and the geometric term in the tangential momentum is

\begin{equation}
S^\mathrm{geom}_{i+1/2,j} = - \check{V}_{i+1/2,j} \frac{ \bar{\rho}_{i+1/2,j}  \bar{u}^r_{i+1/2,j}  u^\theta_{i+1/2,j} }{r_{i+1/2}} .
\end{equation}

\noindent The full semi-discretised formulae are detailed in the appendix \ref{appendix:formulae}.

Spherical coordinates have severe drawbacks in the discretisation of a sphere. First, since $r=0$ is a singular point of the coordinates, the cell size reduces to zero toward the centre. They are therefore inappropriate for modelling the central regions of the star. Also, the cell size increases toward the surface, whereas finer resolution is needed in these regions (see Sect. \ref{giant}). \cite{Calhoun:2008:LRG:1461507.1461512} shows how to design structured grids that overcome these limitations.

\subsection{Temporal method}
\label{temporal}

\subsubsection{CFL time steps}
\label{dtCFL}

When an explicit method is used to integrate a discrete set of equations in time, there is a constraint on the time step that results from the CFL condition. The CFL condition states that the physical domain of dependence of the solution should be included in the numerical domain of dependence, otherwise the numerical scheme cannot be convergent. For an explicit scheme, this yields a constraint on the time step that has to be smaller than the CFL time step to perform a stable computation (independently of any accuracy consideration).

Determining the stability limit for a given explicit scheme is not always trivial, if not impossible, especially in the non-linear and multi-dimensional cases. See, for instance, \cite{Hirsch:1990book2} for a careful derivation of the CFL time step for popular schemes. The CFL time steps that we define in this section are based on very simple arguments and should only be considered as a guideline for the maximum stable time step.

For the hydrodynamic equations, which are hyperbolic equations, the CFL condition states that the time step should not be longer than the crossing time of the fastest wave over a cell. In this case the maximum stable time step is equal to

\begin{equation}
	\label{hydro_cfl}
	\Delta t_\mathrm{hydro} = \min \frac{\Delta x}{|u| + c_s},
\end{equation}

\noindent where $|u| + c_s$ is the velocity of the fastest wave ($u$ is the velocity of the background flow).

When time-dependent diffusion processes are considered, for instance in our case radiative diffusion, the corresponding CFL time step is related to the typical time scale for diffusion in a cell:

\begin{equation}
	\label{parabolic_cfl}
%	\Delta t_\mathrm{rad} = \frac{\Delta x^2}{\max( \chi )}
	\Delta t_\mathrm{rad} = \min \frac{\Delta x^2}{\chi},
\end{equation}

\noindent where $\chi=k_\mathrm{rad}/(\rho c_p)$ is the radiative diffusivity coefficient, and $c_p$ the specific heat at constant pressure. For advection-diffusion problem, which are of mixed nature, the CFL time step is taken to be the minimum values of equations (\ref{hydro_cfl}) and (\ref{parabolic_cfl}).

We also define here a useful time step that we call the \emph{advective time scale}. It is defined by

\begin{equation}
	\label{advective_cfl}
	\Delta t_\mathrm{adv} = \min \frac{\Delta x}{|u|}.	
\end{equation}

\noindent Its definition is inspired by the hydro time step (\ref{hydro_cfl}) where we have set the velocity of the wave to be simply $u$ instead of $u+c_s$. This time step characterises the time needed for the flow to cross one cell. We stress that, for a code solving the standard compressible hydrodynamic equations with an explicit method, this time step is \emph{not} a stability limit. We introduce this time scale since, as shown later, it corresponds to a natural limit for the time step; however, for anelastic codes, this time step represents the stability limit, since sound waves are filtered out in this approach.

With the definition of these three time steps, we can define the corresponding CFL numbers for a given numerical time step $\Delta t$:

\begin{equation}
\mathrm{CFL}_\mathrm{hydro} = \frac{\Delta t}{\Delta t_\mathrm{hydro}}; \ \mathrm{CFL}_\mathrm{rad} = \frac{\Delta t}{\Delta t_\mathrm{rad}};\  \mathrm{CFL}_\mathrm{adv} = \frac{\Delta t}{\Delta t_\mathrm{adv}}.
\end{equation}

By adopting an implicit integration strategy, our goal is to release the stability limit on the time step. Therefore, to appreciate the performance of our solver, we carefully monitor the values of these three CFL numbers.

In 2D, the time steps defined above are computed independently in both directions and the CFL time step is defined as the minimum of these values.

\subsubsection{Implicit strategy}
\label{implicitstrategy}

After the spatial discretisation method described in the previous section, one obtains a system of coupled ODEs (see Appendix \ref{appendix:formulae}):
\begin{equation}
\label{eq:MethodOfLine}
\frac{dU_{i,j}}{dt} = R(U_{k,l}),
\end{equation}

\noindent where $U_{k,l}$ is a vector gathering all the quantities appearing below time derivatives (i.e. $\rho$, $\rho e$, $\rho u_r$, and $\rho u_\theta$, multiplied by the appropriate volumes), and $R(U_{k,l})$ contains all fluxes and sources $s$:

\begin{equation}
\begin{aligned}
R(U_{k,l}) = - ( &S^r_{k+1,l} F^r_{k+1,l} - S^r_{k,l} F^r_{k,l} \\
	+ & S^\theta_{k,l+1} F^\theta_{k,l+1} - S^\theta_{k,l} F^\theta_{k,l}) \\ 
	+ & s_{k,l}V_{k,l}.
\end{aligned}
\end{equation}

To obtain an update formula we now apply an implicit temporal method to integrate (Eq. \ref{eq:MethodOfLine}) in time:

\begin{equation}
\label{eq:nls}
U^{n+1}_{i,j}-U^{n}_{i,j} = \Delta t \big ( \beta R^{n+1} + (1-\beta) R^n \big ) 
\end{equation}

\noindent where $R^n$ (resp. $R^{n+1}$) is the r.h.s. of equation (\ref{eq:MethodOfLine}) evaluated at time $n$ (resp. $n+1$). The first-order accurate Backward Euler method corresponds to $\beta=1$. The second-order accurate Crank-Nicholson corresponds to $\beta=1/2$. Both the Backward Euler and the Crank-Nicholson methods are A-stable, which implies that they are unconditionally stable\footnote{Not all implicit methods are A-stable; e.g. the Adams-Moulton methods are implicit but only conditionally stable.}. For stiff equations, a more relevant property is the L-stability of the scheme \citep[see e.g][]{LeVequeODE}. The Backward Euler method is L-stable but the Crank-Nicholson method is not. This implies that the latter can have difficulties with stiff problems and produce spurious oscillations. Indeed, when computing stellar models with high values of the radiative CFL (see the A-type star model in Sect. \ref{Astar}), we have found that it is more stable to use a value of $\beta$ that is slightly higher than $1/2$, for instance $\beta = 0.55$. In the future, we plan to implement the second-order backward differentiation formula, which is a second-order implicit multistep method having the A \& L-stability property.

Equation (\ref{eq:nls}) describes a system of non-linear equations that defines the solution at the new time step, $U^{n+1}$. We use $\rho$, $e$, $u_r$, and $u_t$ as independent variables, so the quantities that appear in the time derivatives in our equations are thus \emph{composite quantities} of the primitive variables.

\subsubsection{Newton-Raphson procedure}

To solve for the new time step $\rho^{n+1}$, $e^{n+1}$, $u_r^{n+1}$, $u_\theta^{n+1}$, one has to solve the set of non-linear equations $F(U^{n+1})=0$, where the residual $F$ is defined as

\begin{equation}
\label{eq:NonlinearSystem}
F(U^{n+1}) = U^{n+1}-U^{n} - \Delta t \big ( \beta R^{n+1} + (1-\beta) R^n \big ).
\end{equation} 

\noindent Here, we proceed with a Newton-Raphson procedure. The procedure to find $U^{n+1}$ is initialised by taking the current solution as an initial guess for the new solution: 

\begin{equation}
U^{(0)} = U^{n}.
\end{equation}

\noindent At each Newton-Raphson iteration we solve for the corrections $\delta U^{(k)}$:

\begin{equation}
\label{eq:LinearSystem}
J \times \delta U^{(k)} = -  F( U^{(k)}),
\end{equation}

\noindent where $J$ is the Jacobian matrix of the residual vector $F$. The current iteration is then updated with

\begin{equation}
 U^{(k+1)} = U^{(k)} + \lambda \delta U^{(k)},
\end{equation}

\noindent where $0< \lambda \le 1$ is introduced to prevent divergence of the procedure. The classical Newton-Raphson update corresponds to $\lambda=1$. A straightforward line-searching algorithm is applied to find the value of $\lambda$ that ensures a decrease in the residual. However, we found that when solving the non-linear radiative diffusion term at very high radiative CFL numbers ($\sim 10^6$), the first Newton-Raphson iteration always increases the residual. Different tests have shown that this seems unavoidable, and since the system would eventually converge within a few iterations, we always enforce $\lambda=1$ during the first iteration in this particular case.

We stop the iteration procedure when the relative correction drops below a certain level:

\begin{equation}
\left \| \frac{\delta U^{k}}{U_0} \right \|_\infty < \epsilon,
\end{equation}

\noindent where $U_0$ are typical values of the variables. For the density and internal energy we simply choose the corresponding values at iteration $k$, but for the velocities we choose $\max(u, c_s)$, where $u$ is the value of the velocity at iteration $k$, and $c_s$ is the sound speed. In our computations we set $\epsilon=10^{-6}$.

\subsubsection{Jacobian matrix computation}

The Jacobian matrix elements are the partial derivatives of the numerical scheme equations (\ref{eq:NonlinearSystem}) with respect to all numerical variables. We compute the partial derivatives by finite differencing. The location of non-zero elements of the Jacobian depends exactly on the stencil of the numerical scheme (i.e. which variables contribute to a given equation). It is therefore easy to determine the non-zero structure (also known as the sparsity pattern) of the Jacobian matrix. Any Jacobian computation technique should then take advantage of the non-zero pattern to compute the Jacobian matrix elements. We use here the \emph{Coloured Finite Differencing} technique \citep[CFD hereafter; see][]{CURTIS01021974, Gebremedhin05whatcolor}, which minimises the number of function evaluation needed to compute the Jacobian matrix. The CFD algorithm uses the non-zero pattern of the Jacobian to group independent columns (i.e. variables) of the Jacobian matrix, i.e. those that do not share a non-zero element on the same line, into ``colours". This yields a ``column compressed" representation of the Jacobian. The number of colours $n_g$, which is also the number of columns of the compressed Jacobian matrix, is roughly equal to the maximum number of non-zero elements found in a row. This number depends \emph{only} on the stencil of the scheme and is (ideally) \emph{independent} of the dimension of the Jacobian matrix. To compute the matrix elements, the CFD algorithm perturbs all variables associated with the same colour and recomputes the non-linear equations. The matrix elements are then obtained by a straightforward finite difference against a reference value of the equations. The number of non-linear function evaluations is therefore equal to $N(n_g +1)$, where $N$ is the size of the matrix. As $n_g$ is independent from $N$, the CFD algorithm is expected to scale as $N$.

\begin{figure}[t] %  figure placement: here, top, bottom, or page
  \vspace{0.5cm}
   \centering
   \includegraphics[width=6cm]{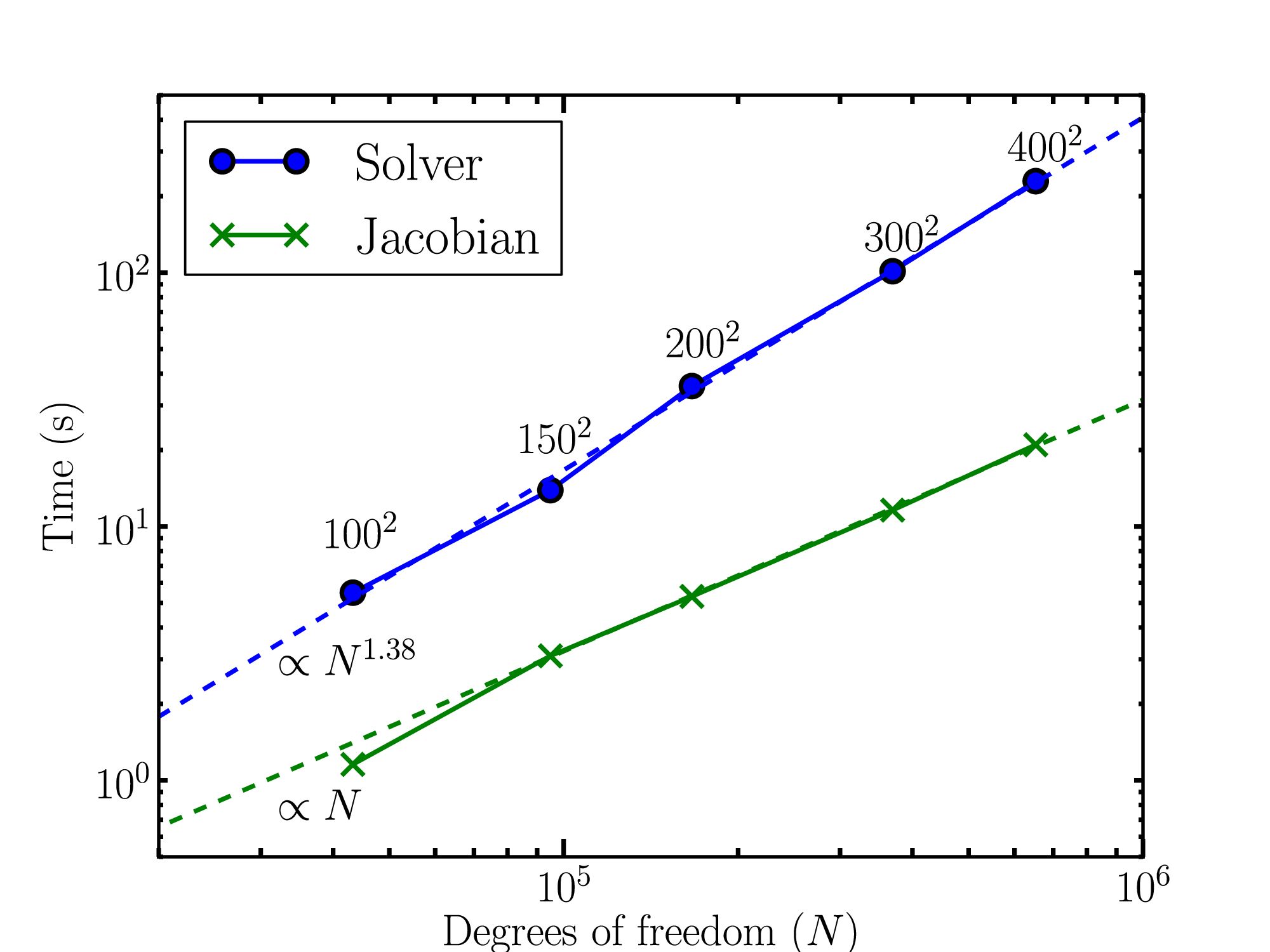} 
      \vspace{0.5cm}
   \caption{Time spent in computing the Jacobian matrix (crosses) and in solving one linear system with MUMPS (dots) for different matrix sizes on a Pentium III Xeon quad-core CPU at 3 GHz.}
   \label{fig:timing}
\end{figure}

In our code, we use the CFD algorithm that is implemented in the Trilinos library (see \citealt{Trilinos-Overview}). The graph-colouring algorithm is the ``greedy" algorithm proposed in \cite{CURTIS01021974}. We find that for our matrices this algorithm typically produces a number of colour $n_G = 51 \pm 1$ for matrix sizes ranging from 2304 ($20^2$ domain) to 652\,864 ($400^2$ domain).  Therefore, the number of colours is roughly independent of the matrix size, and the algorithm scales as $N$. This is illustrated in Fig. \ref{fig:timing}, where we plot the time needed to compute the Jacobian matrix for different spatial resolutions. 

\subsubsection{Linear solver}

The Newton-Raphson procedure requires the solution of the linear equation (\ref{eq:LinearSystem}). Equation (\ref{eq:LinearSystem}) is a sparse linear problem that can be solved by either a direct method (i.e. LU decomposition) or an iterative method. Iterative solvers (e.g. GMRES, BiCGStab) perform usually faster than direct methods but they rely heavily on preconditioning to improve (or even obtain) convergence. On the other hand, direct solvers are more accurate and robust but they become very expensive as the size of the linear system increases. For convenience, and since our 2D approach allows it, we decided to use a direct solver for developing the code. We use in our code the MUMPS solver (MUltifrontal Massively Parallel Solver, see \citealt{amestoy:15,Amestoy06hybridscheduling}). MUMPS is able to handle general non-symmetric matrices; i.e., it is not restricted to tridiagonal or pentadiagonal matrices. MUMPS achieve robustness by detecting null pivots during the elimination phase and accuracy by applying a few (usually 2 or 3) steps of iterative refinement in the postprocessing phase. %MUMPS optionally provides backward error analysis of the solution, which puts a constrain on the accuracy of the provided solution (GIVE NUMBERS).

\begin{figure*}[t] %  figure placement: here, top, bottom, or page
  \centering
  \parbox{0.49\linewidth}{\includegraphics[width=0.65\linewidth,viewport=20 20 500 620,angle=90,clip]{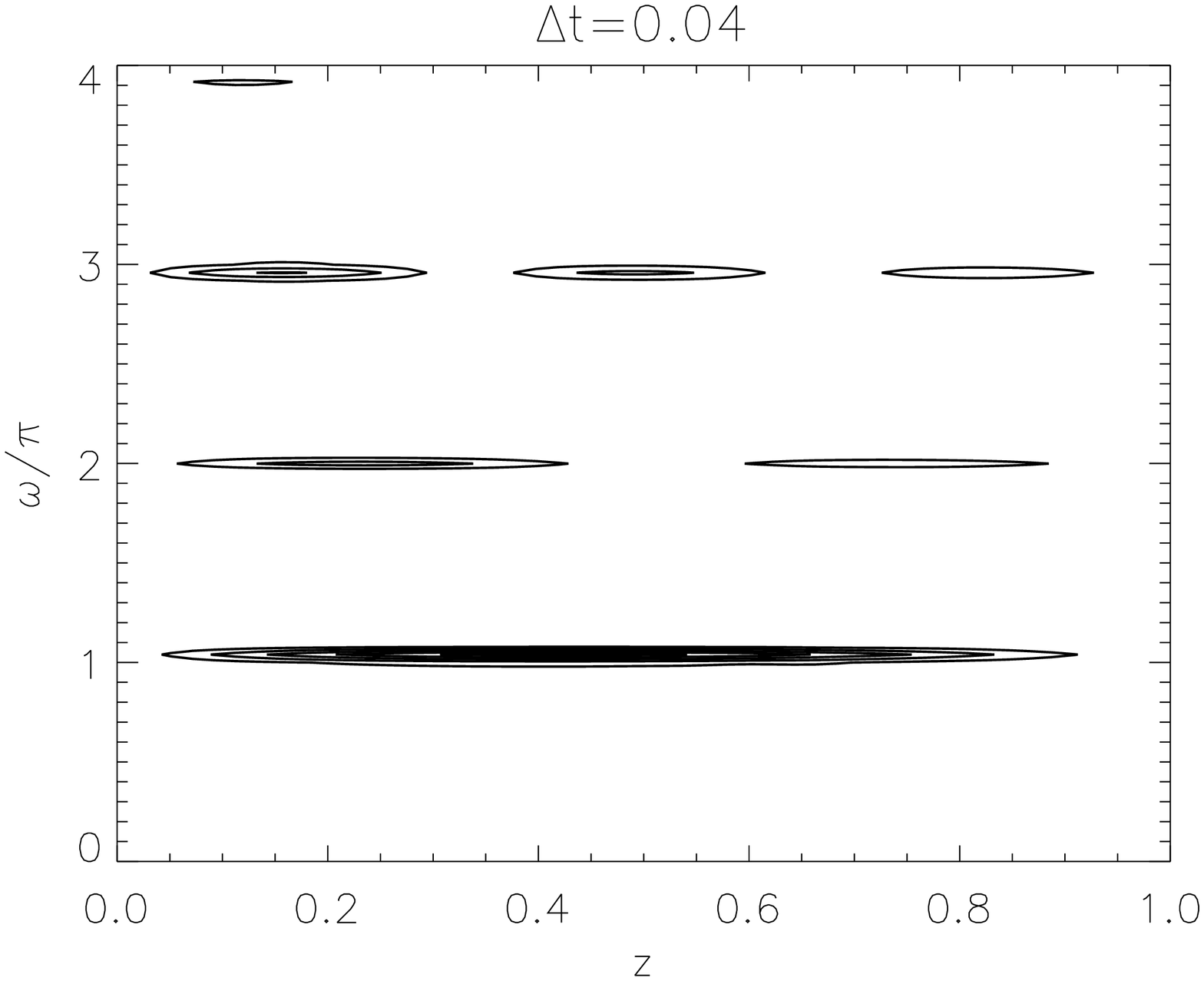}}
  \parbox{0.49\linewidth}{\includegraphics[width=0.65\linewidth,viewport=20 20 500 620, angle=90,clip]{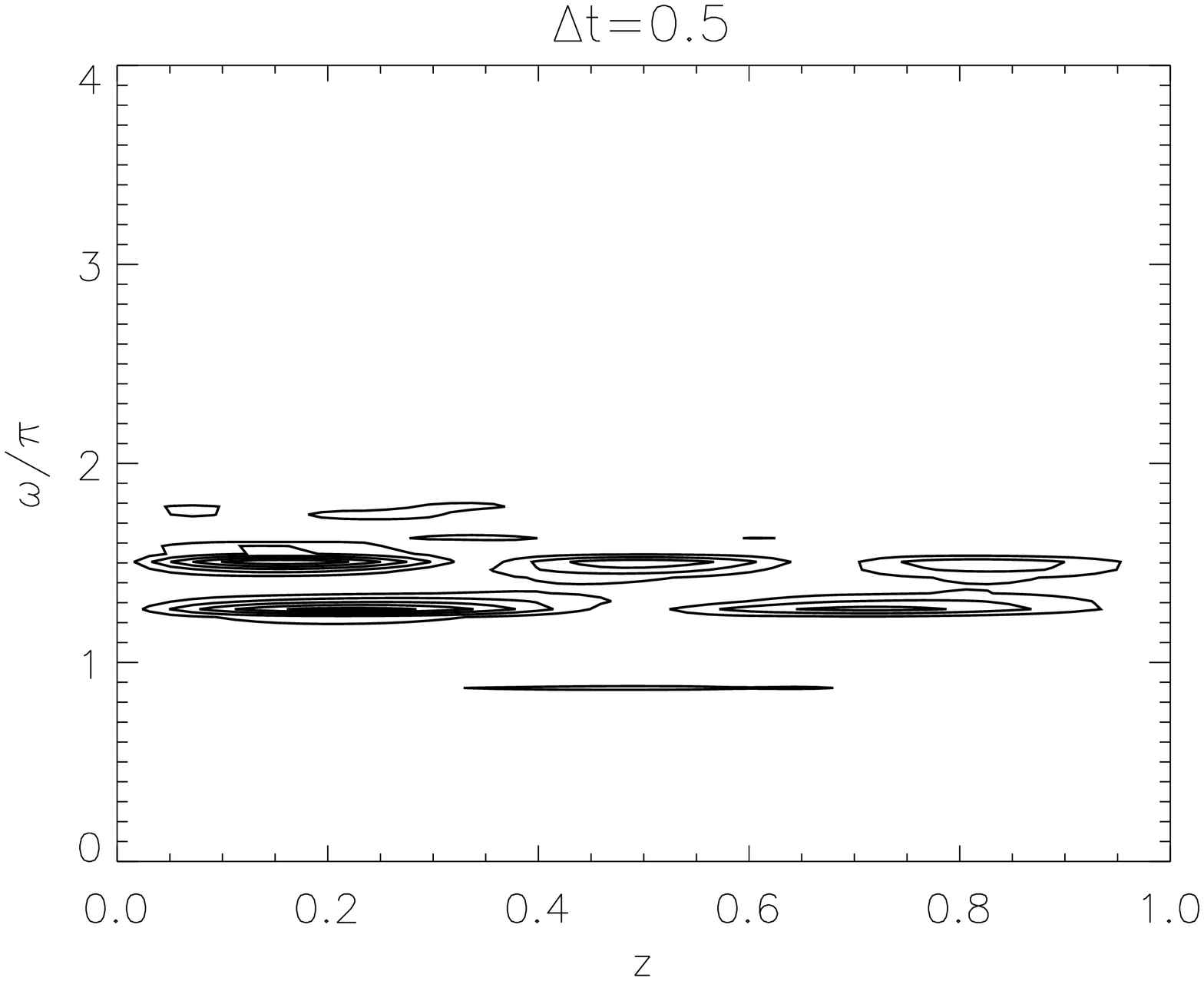}}
   \caption{Accoustic modes in the $(z,\omega)$ plane for $\Delta t = 0.04$ (left) and $\Delta t = 0.5$ (right).}
   \label{fig:bubble_acc_cfl}
\end{figure*}

We use the advantage of knowing the non-zero structure of the Jacobian matrix to perform the symbolic factorisation of the linear system only one time, at the beginning of a simulation. Later on, only the numerical factorisation is performed with a given r.h.s. vector to obtain the solution of the linear system. The cost of the direct solver is illustrated in Fig.~\ref{fig:timing} where the time needed to solve one linear system is shown for various spatial resolutions. The linear solver dominates the time needed to compute the Jacobian matrix. With a simple linear regression in the log-log space, we find that in our case (i.e. for the Jacobian matrices that result from our temporal and spatial discretisation) the time needed to solve one linear system scales as $N^{1.38}$. For 2D runs with resolution up to $500^2$, the number of variables is under one million. Linear systems are in this case still tractable with direct methods. The memory requirement for the LU decomposition ranges from 10 Mb for 2304 degrees of freedom to 9.4 Gb for 652\,864 degrees of freedom. It is clear that larger systems (e.g. 3D problems) will require too much memory and computational time to be tractable. Therefore, we plan in the future to implement iterative solvers that are less expensive both in terms of memory and CPU time.

\subsubsection{Time step control}

In an implicit method, one has to introduce a time step control strategy, usually based on empirical grounds. We found that the following three strategies are useful, in the order of the most to the least basic:

\begin{description}
\item[\bf{Convergence-based}:] If the Newton-Raphson procedure converges within five iterations, increase the time step by a factor of 1.5. If NR converges in more than ten iterations, decrease the time step by a factor of 2.
\item[{\bf Relative change-based}:] Compute relative change in the solution between time step $n$ and $n+1$. If it is lower than 10 \%, increase the time step by a factor of 1.5. If it is larger than 20 \%, decrease time step by a factor of 2.
\item[{\bf Advection-based}:] Set the time step so that CFL$_\mathrm{adv}=1$.
\end{description}

The last strategy is physically motivated since it ensures that the fluid does not move across more than one cell width during one time step. For the preliminary results presented in this paper, we have used the first, least stringent, and least physical, strategy. In the future we plan to compare these three strategies, both in terms of accuracy and computational time. Finally, in some cases, we also provide maximum/minimum values for the CFL number or the time step.

\section{Test: oscillations of an entropy bubble}
\label{section:tests}

\begin{table}[t]
\caption{Run parameters for the entropy bubble test.}             % title of Table
\label{table:bubble}      % is used to refer this table in the text
\centering                          % used for centering table
\begin{tabular}{c c c c c}        % centered columns (4 columns)
\hline\hline                 % inserts double horizontal lines
Run & Final time & Resolution & time step & CFL \\    % table heading 
\hline                        % inserts single horizontal line
   1 & 200 & $50^2$ & 0.04 & 1 \\      % inserting body of the table
   2 & 200 & $50^2$ & 0.5 & 12.5 \\
%   3 & 50 & $150^2$ & 0.04 & 3 \\
%   3 & 200 & $150^2$ & 0.5 & 37.5 \\
\hline                                   %inserts single line
\end{tabular}
\end{table}

Basic tests (see appendix \ref{appendix:tests}) have been successfully done for code validation during its first stages of development, so we focus here on a test that is more relevant to the physical processes we would like to investigate. It was taken from \citealt{2004A&A...421..775D} (DB hereafter). We considered an isothermal stratified atmosphere with constant gravity perturbed by an entropy bubble. We computed the oscillations of this bubble around its equilibrium position, and we analysed the spectrum of internal gravity and sound waves driven by the bubble. The initial setup, parameters, and units normalisation are similar to DB and are therefore not reproduced here. The Brunt-V\"ais\"al\"a frequency of the atmosphere is $N\sim 0.82$. For this test, Eq. (1-3) were solved with the viscous terms included. The kinematic viscosity coefficient $\nu$ was set to $5\times10^{-4}$ in all runs. The thermal conductivity is set to zero, as we were interested in the adiabatic evolution of the system. We used a Cartesian version of the code\footnote{Cartesian coordinates are used in this particular case because of analytical solutions in this geometry.} and considered a domain $(x,z) \in [-0.5,0.5]\times[0,1]$. Periodic boundary conditions were used in the horizontal directions and non-penetrative, stress-free conditions were used at the bottom and top boundaries. All the run parameters are summarised in Table \ref{table:bubble}. For each run we set the time step to a constant value.

As in DB, we describe each wave mode by two integers $l,n \ge 0$, which are the number of nodes in the horizontal and vertical directions, respectively. In this test we looked at two types of waves: 

\begin{enumerate}
\item Vertical sound waves that have frequencies $\omega = (n+1)\pi$. These waves have $l=0$ since they have no horizontal structure. 
\item Internal gravity waves that have frequencies $\omega < N$ and are characterised by $l\ne0$ (internal gravity modes cannot propagate in the vertical direction).
\end{enumerate}

See DB for the derivation of these properties. These two kinds of waves occupy well-separated regions of the frequency spectrum. Vertical sound waves have vertical periods $P \le 2$ and will therefore be more sensitive to the time step. Internal gravity modes have longer periods $P \gtrsim 7.3$ and will therefore be less sensitive to the time step. This test thus provides a good analysis of the influence of the time step on the accuracy of the numerical results.

We first analysed the vertical sound wave spectrum by using method 2 described in DB, for runs 1 \& 2 (cf. Table \ref{table:bubble}). In this method, the wave spectrum is represented in the $(z,\omega)$ plane. For any given time step $\Delta t$, the Nyquist theorem states that no periodic signal with frequency higher than $\omega_N = \pi / \Delta t$ can be resolved. Also the temporal Fourier transform has a frequency resolution that is $\Delta \omega = 2\pi/T$, where $T$ is the duration of the simulation. The results are shown in Fig. \ref{fig:bubble_acc_cfl} for two different time steps. For $\Delta t = 0.04$, we recognise the signature of the $n=0,1,2$ modes. The $n=3$ mode is below the range of the isocontour levels. Furthermore, all the modes are located at the correct frequencies (within $\Delta \omega \simeq 0.125$). For $\Delta t = 0.5$, one has $\omega_N = 2\pi$ thus normally allowing only for modes $n=0$ and $n=1$. In the map, one can recognise the signature of the $n=0,1,2$ modes. None of them is located at the correct frequency, and it is surprising to find the $n=2$ mode whose eigenfrequency is above the Nyquist frequency. In this case it is clear that the time step is too large to accurately resolve the sound waves. To analyse the gravity modes, we used the method developed in DB: we projected the velocity field on the anelastic eigenvectors. With this method we obtained the individual mode amplitudes $c_{ln}$ which are the coefficients in the eigenvector expansion. With such a decomposition it is straightforward to obtain the mode frequency. We checked that, for both runs, the gravity mode frequencies match the analytical predictions to the percent level.

% Requires the booktabs if the memoir class is not being used
\begin{table*}[t]
\caption{Stellar parameters of the models considered in this paper.}
\label{tab:StellarParameters}
\centering                          % used for centering table
\begin{tabular}{c c c c c c c}        % centered columns (4 columns)
\hline\hline                 % inserts double horizontal lines
   Model   & Mass ($M_\odot$) & R ($R_\odot$) &$T_\mathrm{eff}$ (K) & $\log(g)$ & Luminosity ($L_\odot$) & Comments \\
\hline                        % inserts single horizontal line
%        I   & 20 & &70\,000 & & $10^5$ & Z=0\\
       I   & 2 & 2.1 &8\,500 & 4.10 & $20$ & A-type star\\  
      II   & 5 & 58.8 &4\,500 & 1.6 & $ 1,258$ & Cold giant\\
\hline                         %inserts single line
\end{tabular}
\end{table*}

This test highlights the importance of the choice of the time step to correctly describe the physical process of interest. In the present case, sound-wave amplitudes are much lower than the amplitudes of the gravity modes (the initial setup is almost in hydrostatic equilibrium), and therefore they do not play an important role in this problem. One can then safely use a higher time step (e.g. $\Delta t=0.5$) to study the internal gravity modes, as shown by the excellent agreement with predicted frequencies. Obviously, one could not do that if the initial setup was intended to be far from hydrostatic equilibrium, since sound waves would at least be as important as internal gravity modes.

\section{Stellar models}
\label{section:stars}

We now test our code on realistic stellar conditions.

\subsection{Local simulation: A-type star}
\label{Astar}

\begin{table*}[tb]
\caption{Numerical parameters for the A-star run.}
\label{table:Astar}      % is used to refer this table in the text
\centering                          % used for centering table
\begin{tabular}{c c c c c c c c c}        % centered columns (4 columns)
\hline\hline                 % inserts double horizontal lines
Resolution & Domain & $\tau_\mathrm{KH}$ & $\Delta t_\mathrm{hydro}^\mathrm{CFL}$ & $\Delta t_\mathrm{rad}^\mathrm{CFL}$ & Final time & Time steps & Wall Time & Mem.\\
\hline                        % inserts single horizontal line
$214\times 512$ & $[0.974,1.]\times[88.2^\circ, 91.8^\circ]$ & 1.15 d & 1.84 s & $3\times 10^{-2}$ s & 3.5 d & 4032 & 2 w 5 d &  4.8 Gb\\
\hline                         %inserts single line
\end{tabular}
\end{table*}

\begin{figure}[t] %  figure placement: here, top, bottom, or page
   \centering
   \vspace{0.5cm}
   \includegraphics[width=7cm]{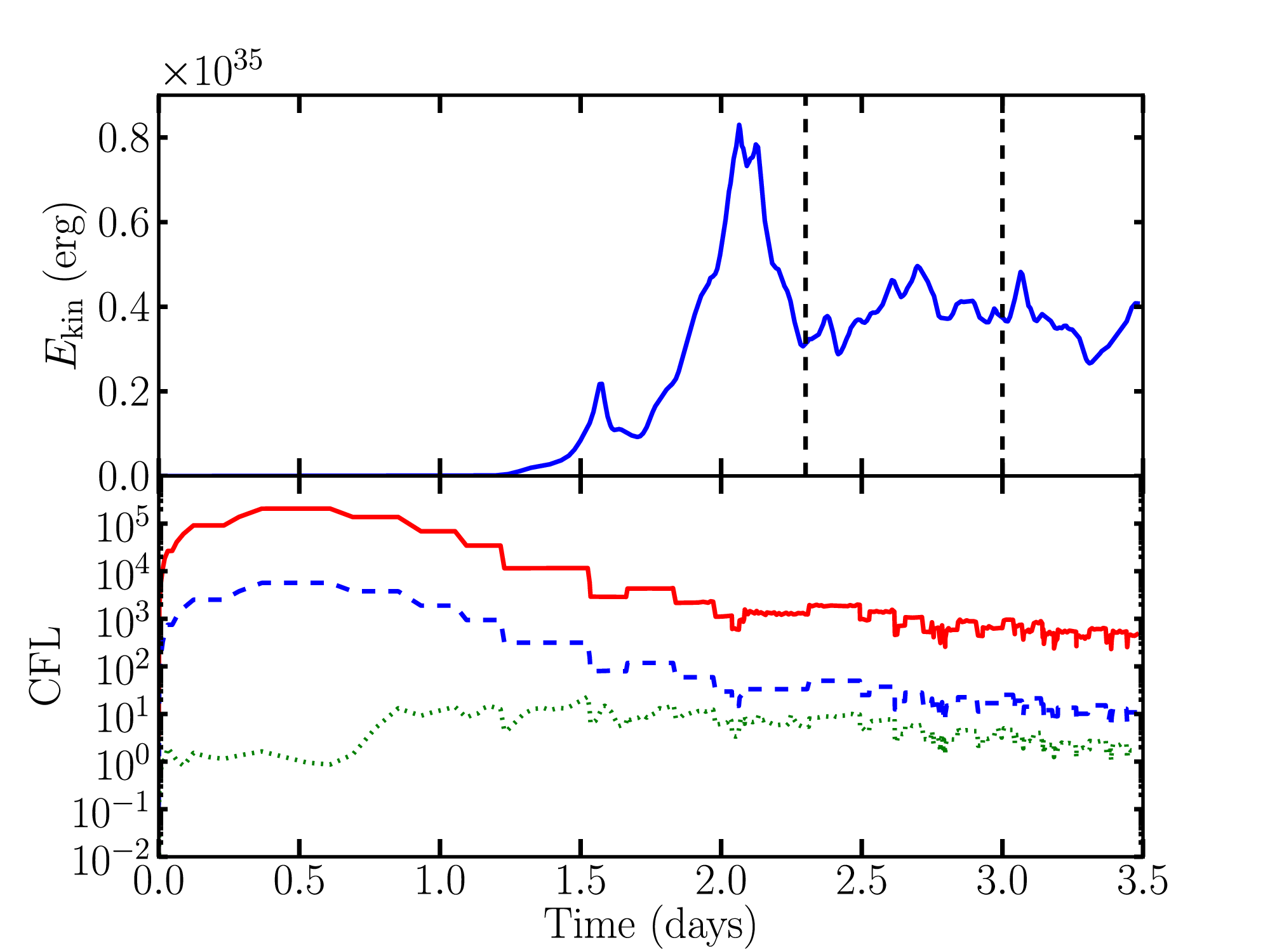} 
   \vspace{0.5cm}
   \caption{Evolution of the total kinetic energy (top panel) and CFL numbers (lower panel, solid: radiative; dashed: hydro; dotted: advection) for the A-type star run. The vertical dashed lines show the time interval on which the averaged fluxes shown in Fig. \ref{fig:run1_Fconv} are computed.}
   \label{fig:run1_Ekin_CFL}
\end{figure}

\begin{figure*}[t] %  figure placement: here, top, bottom, or page
   \centering
   \includegraphics[angle=90,width=0.9\linewidth,viewport=45 30 350 680, clip]{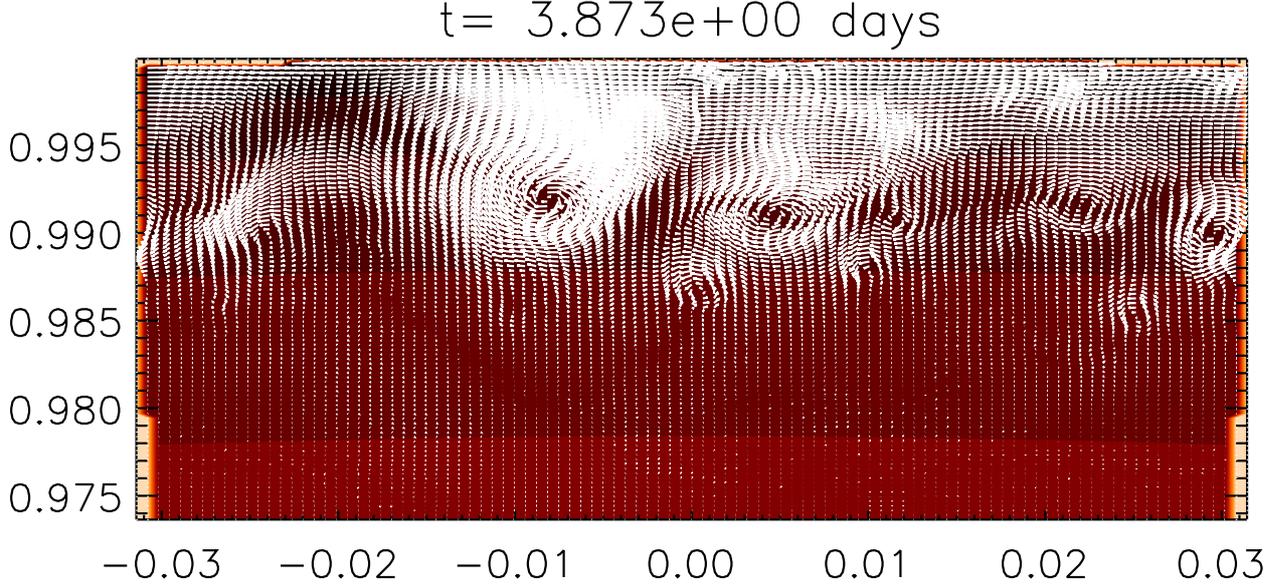} 
   \caption{Snapshot of our simulation of an A-type star. The axes are Cartesian coordinates normalised by the stellar radius and with the origin at the star centre.}
   \label{fig:Astar_SN}
\end{figure*}

We present in this section results of convection in A-type stars (see model I in Table \ref{tab:StellarParameters}), since comparison with previous models of such stars, calculated by different groups using explicit codes (see \citealt{1984ApJ...282..550S,1996A&A...313..497F}), are possible. These types of stars are weakly convective and possess two convectively unstable zone near their surface: around $T \sim 40\,000$ K due to HeII ionisation and near $T\sim 10\,000 $ K due to HeI/H ionisation. Usually the HeI/H surface convective zone is strong, but very narrow, whereas the HeII convective zone is broader but much weaker. To model these convective regions it is therefore sufficient to focus on the surface layers of the star by performing \emph{local} simulations.

Our initial model starts at the photosphere, i.e. the location where $\tau=2/3$ and $\Teff=8\,500$ K. This particular choice inhibits the convection in the HeI/H ionisation zone, which is located at the photosphere, so we do not expect to model this convective region correctly. A better modelling of this region should include a part of the stellar atmosphere in the computational domain, as is done for instance by \cite{1996A&A...313..497F}. Our initial models are produced with a 1D stellar structure code \citep[see][]{1991A&A...245..548B} where the mixing length theory (MLT hereafter) prescription for convection has been turned off. The resulting fully radiative models are very close to the MLT models (even near the surface) since convection is too weak, in terms of energy transport,  to significantly modify the structure. The model presented here extends down to a depth where the temperature is $T=100\,000$ K, where the star is radiatively stable, so that the numerical domain includes a stable, radiative, buffer zone at the bottom of the domain. Since our simulation focusses on a very small fraction of the star, the geometry is Cartesian-like (but we use spherical geometry). In terms of physical dimensions, the physical domain has an average width of $90\,000$ km and a height of $37\,500$ km.

To construct the initial model for the multi-dimensional code, we simply copy the 1D structure along the tangential direction. However, the number of tangential grid points has to be carefully chosen  in order to keep an acceptable aspect ratio of the cells. The number of grid points in the tangential direction therefore depends on the number of grid points in the radial direction through the radial extend of the cells.

We discuss below the results of our run. Table \ref{table:Astar} summarises important parameters. $\tau_\mathrm{KH}$ is the thermal time scale defined as the ratio between the total internal energy in the computational domain divided by the luminosity of the model. The last column of Table \ref{table:Astar} shows the memory requirement for the LU decomposition. Calculations were done on an AMD ``Istanbul" processor (6 cores at 2.7 Ghz). Figure \ref{fig:run1_Ekin_CFL} shows the evolution as a function of time of the total kinetic energy in the domain (used to track the development and saturation of the convective instability) and the evolution of the different CFL numbers defined in Sect. \ref{dtCFL}. At the beginning, the model undergoes a relaxation toward hydrostatic equilibrium, and the CFL number rapidly reaches values $>$ $10^5$. Eventually, the velocity  field that develops as a result of the relaxation process triggers the convective instability, which corresponds to an increase in the kinetic energy to a significant level. As the first vortices develop, the time step decreases since the Newton-Raphson procedure would not converge otherwise. After $\sim 2.3$ days of simulated time, we reach a quasi-steady state that lasts until the end of the simulation at $t \sim 3.5$ days. In this phase, the time step remains roughly constant with $\Delta t = 18 \pm 2.6$ s, corresponding to CFL$_\mathrm{hydro}=9.8 \pm 1.4$ and CFL$_\mathrm{rad}=400 \pm 60$ (the second value indicated correspond to the standard mean deviation). In this quasi steady state, convection is fully developed, and we observe a complex vortex dynamics involving vortex pairing/merging and the formation of downdrafts. These phenomena are characteristic of turbulent convection (see snapshot in Fig. \ref{fig:Astar_SN}).

  \begin{figure}[t] %  figure placement: here, top, bottom, or page
   \centering
   \includegraphics[width=\linewidth]{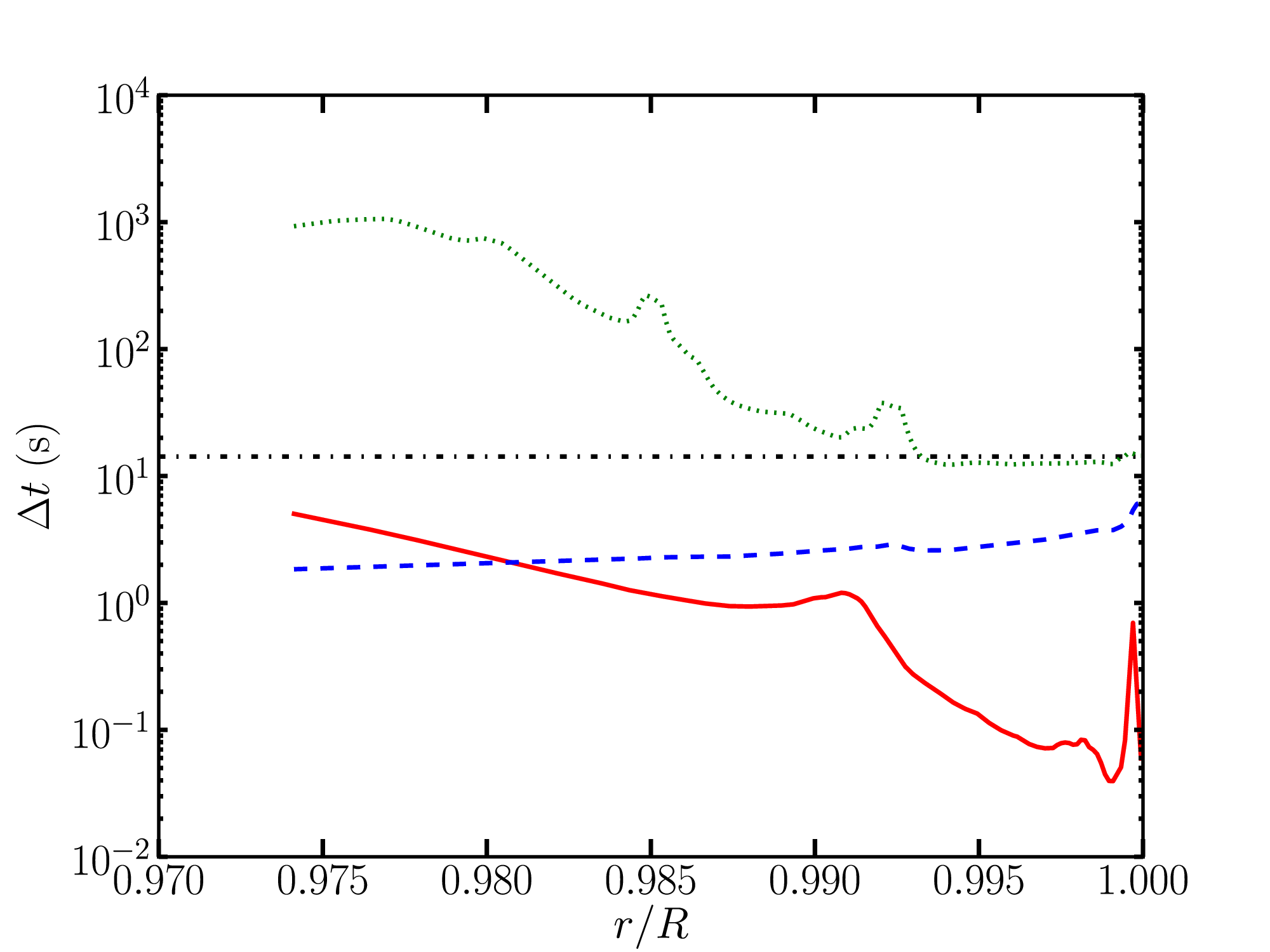} 
   \caption{Radial profiles of the CFL time steps for the A-type star run (see text). From bottom to top: radiation (solid red), hydro (dash blue), and advection (dot green). The horizontal dash-dotted line indicates the time step of the snapshot under consideration.}
   \label{fig:Astar_dt}
\end{figure}

Figure \ref{fig:Astar_dt} shows the radial profiles of the different CFL time steps defined in Sect. \ref{dtCFL}. The CFL time steps are first computed in each cell of the domain, and the radial profiles shown in Fig. \ref{fig:Astar_dt} are then deduced by taking the \emph{minimum} value along the tangential direction for each radius. We first note that the radiative CFL time step provides the most stringent constraint. Solving the radiative diffusion with an explicit method would require a time step lower than $\Delta t = 3\times 10^{-2}$ s and would make any calculation cumbersome. The hydro CFL time step increases with radius, so that the hydro limitation on the time step would come from the bottom of the numerical domain. This is expected since we are using a uniform grid, and at the bottom of the envelope, the temperature, and hence the velocity of the sound waves, are the largest. The advective time scale shows the region where the flow velocities are the largest, i.e. near the surface. The time step of the actual simulation clearly corresponds to the lowest value of the advective time scale. If the time step was much longer than the advective time scale, the solution would change significantly since eddies would be moving across more than one grid cell. In this case, however, the Newton-Raphson procedure would have convergence problems since the initial guess (the solution at time step $n$) would strongly depart from the correct solution. Since our time step strategy is based on limiting the number of iterations, the time step would be automatically decreased. Therefore, we do not expect our current implicit method to work with much larger time steps, and we identify the minimum value of the advective time scale as a natural (physical) limit for the time step.

\begin{figure}[t] %  figure placement: here, top, bottom, or page
   \centering
   \vspace{0.5cm}
   \includegraphics[width=7cm]{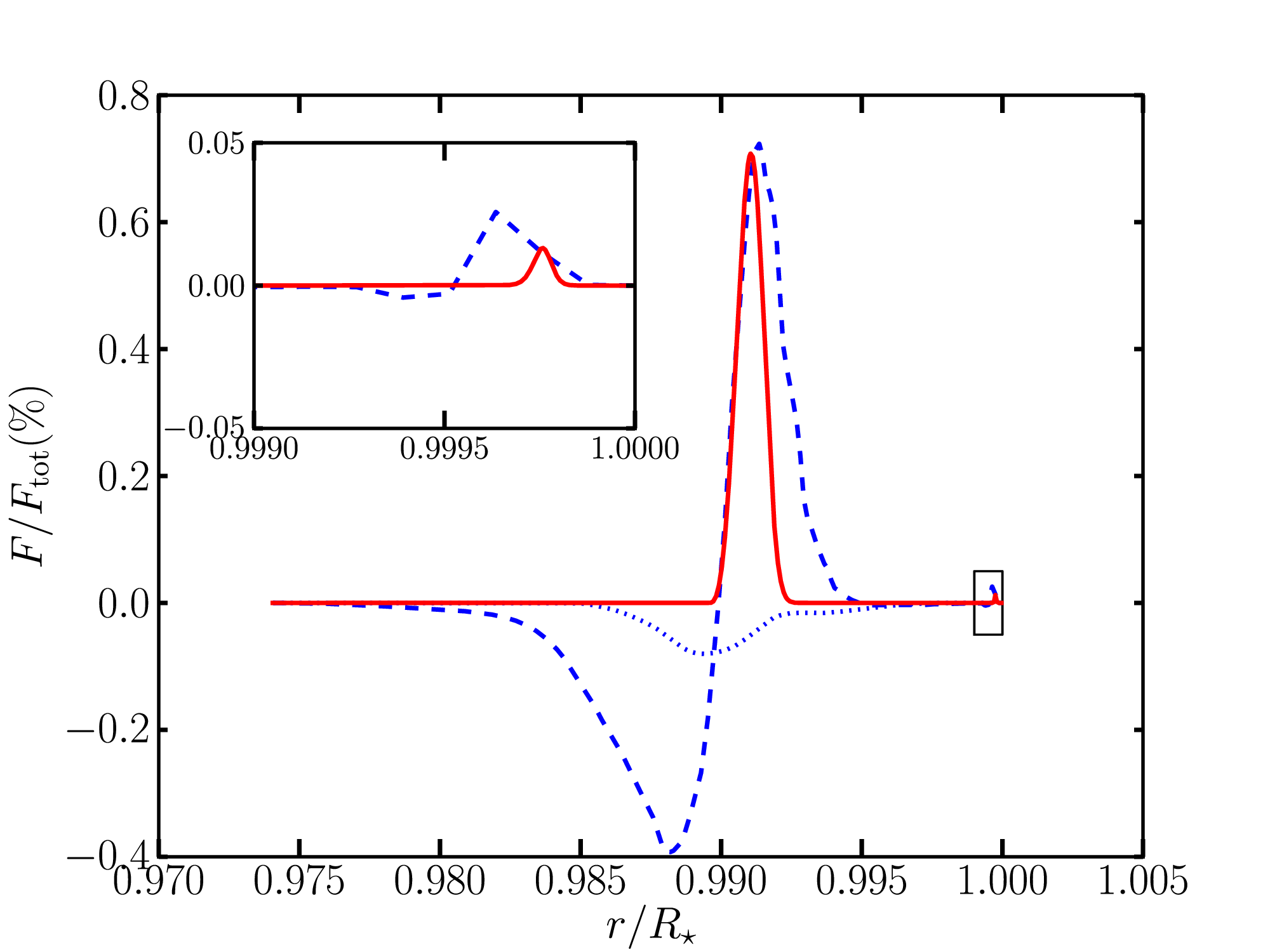} 
    \vspace{0.5cm}   
   \caption{Radial profiles of averaged fluxes for the A-type star run (dashed line: enthalpy flux, dotted line: kinetic energy flux). The solid line is the convective flux as predicted by the mixing length theory with a mixing length parameter $\alpha=1.375$. The box in the upper left corner is a zoom of the surface layer of the star.}
   \label{fig:run1_Fconv}
\end{figure}

We now analyse the properties of the fully developed convective state. We computed the average of the radiative, mass, kinetic, and enthalpy fluxes: 

\begin{eqnarray*}
F_\mathrm{rad} &=& \langle \langle S^r (-k \frac{dT}{dr})\rangle_\theta \rangle_t \\
F_\mathrm{m} &=&  \langle \langle S^r \rho u_r \rangle_\theta \rangle_t\\
F_\mathrm{kin} &=&  \langle \langle S^r u^r \frac{1}{2} \rho v^2\rangle_\theta \rangle_t\\
F_\mathrm{enth} &=& \langle \langle S^r u^r (\rho e + P)\rangle_\theta  - \langle(\rho e + P)\rangle_\theta F_\mathrm{m}\rangle_t.
\end{eqnarray*}.

These quantities have the dimension of a flux multiplied by a surface (i.e. a luminosity), in order to take geometrical effects introduced by the spherical coordinates into account. We however use the term ``flux" in the following to make the discussion clearer. The total flux is defined as the sum of the enthalpy, radiative, and kinetic energy fluxes.

\begin{figure*}[t] %  figure placement: here, top, bottom, or page
   \centering
   \includegraphics[width=0.7\linewidth]{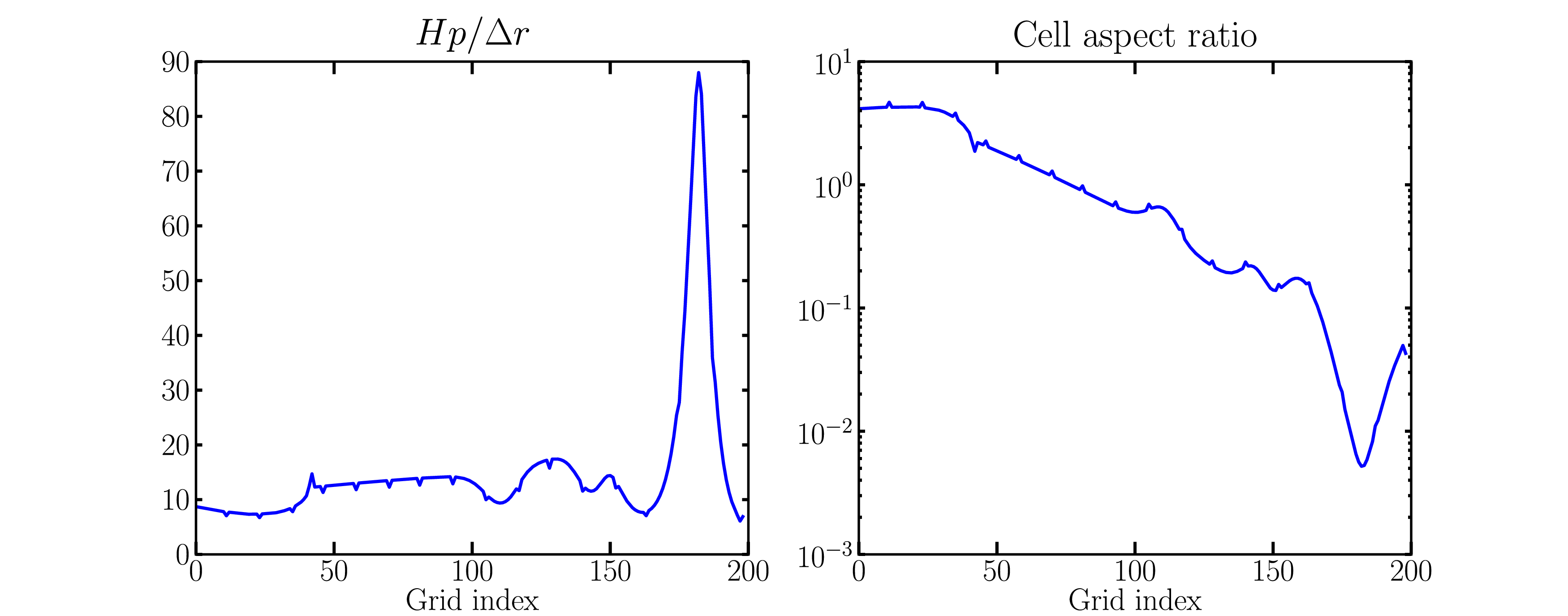} 
   \caption{Properties of the sampled grid for the giant model. Left panel: Pressure scale height to radial cell size ratio. Right panel: Cell aspect ratio when the radial mesh is extended in the tangential direction.}
   \label{fig:Cold_grid1}
\end{figure*}

Figure \ref{fig:run1_Fconv} shows the radial profile of the averaged fluxes defined above. The averaging was done on the interval 2.3 days - 3 days (represents 1408 time steps) but we checked that, by taking smaller windows (but still large enough to have several hundreds of time step), we would obtain similar results. Figure \ref{fig:run1_Fconv} shows the typical profile of the enthalpy flux in the HeII convective zone, as expected from general properties of convection: an inward/outward directed flux, with an outward part stronger than the inward flux. Also the kinetic energy flux is typical of what is expected from convection: downward, as a result of the acceleration of the downdrafts by gravity. The outward convective flux reaches a maximum amplitude of about 0.68 \% of the stellar flux. Depending on the time interval on which the averaging was done, this maximum value ranges between 0.6 \% and 0.77 \% of the stellar flux. This corresponds to the prediction of MLT for an $\alpha$ parameter in the range $1.35 -1.4$. As discussed above, we do not expect to have realistic results for the H/HeI top convective zone since our outer boundary condition at the photosphere inhibits convection in this region. However, despite the lack of resolution (only a few grid points resolve the ionisation region), we still find a non zero average convective flux  with an amplitude of $\sim 10^{-4} F_\star$. For the aforementioned value of $\alpha$, tuned to match the HeII zone, the MLT predicts a convective flux in the HeI/H region, which has a location and an amplitude consistent with our results.

We now compare our results with the work of \cite{1984ApJ...282..550S} and \cite{1996A&A...313..497F}, who have performed numerical models of a A-type star with the same  parameters as ours  (model A5 of Sofia \& Chan paper, model AT85g41N3 of Freytag et al. paper). Sofia \& Chen also considered models extending to the photosphere, but they did not find the small convective flux that appears near the surface in our simulation, most certainly because of the lack of resolution in their computations (their resolution is twice lower than our). Since only \cite{1996A&A...313..497F} resolve the HeI/H region properly, we restrict the comparison to the HeII convective zone. For this region, \cite{1984ApJ...282..550S} found a convective flux between 0.15 and 0.2 \% of the stellar flux (depending on the aspect ratio of their model) and \cite{1996A&A...313..497F} found a convective flux about $0.12 \%$ of the stellar flux. This is significantly lower than the amplitude deduced from our simulations. This difference likely results from different input physics (metallicity, EOS, opacities).

Our results on A-type stars show that we could successfully describe 2D convective patterns in realistic stellar conditions with our fully implicit scheme. Thanks to this approach, the time step in our computation is not limited by sound waves or by radiative diffusion. However, we have shown that the time step is limited by the velocity of the fluid itself, which determines the rate of change in the solution. The hydro CFL cannot be much larger than $10$ when convective motions are fully developed. On the other hand, the radiative CFL number is very high ($ \sim 400$). This is expected for the local models of ``hot" stars that have very short radiative time scales in the low-density, photospheric region. The rather low value of the hydro CFL number and the high value of the radiative CFL number suggest that, for this type of local model, a mixed approach (explicit hydrodynamic + implicit radiation) could allow substantial speedup to be reached as compared to a fully explicit/implicit approach. However, our goal with the present simulation of A-type stars is not to compete with previous simulations performed with explicit codes, which are doing an excellent job for this type of star with very tiny atmospheric and subphotospheric convective zones. Also a realistic description of convection requires a 3D approach. The present study is meant  to test the behaviour and the reliability of the implicit code to model convection under realistic conditions that have been already studied. Nevertheless, we stress that a fully implicit scheme remains very efficient at computing the initial development of convective instability, since the time step can grow to high values as long as velocities remain low.

\subsection{Global simulation: cold giant star}
\label{giant}

The next step is to compute models involving a significant fraction of the star. We consider here a model with parameters corresponding to a cold giant star (model II in Table~\ref{tab:StellarParameters}). 

\begin{figure*}[t] %  figure placement: here, top, bottom, or page
   \centering
   \includegraphics[width=0.7\linewidth]{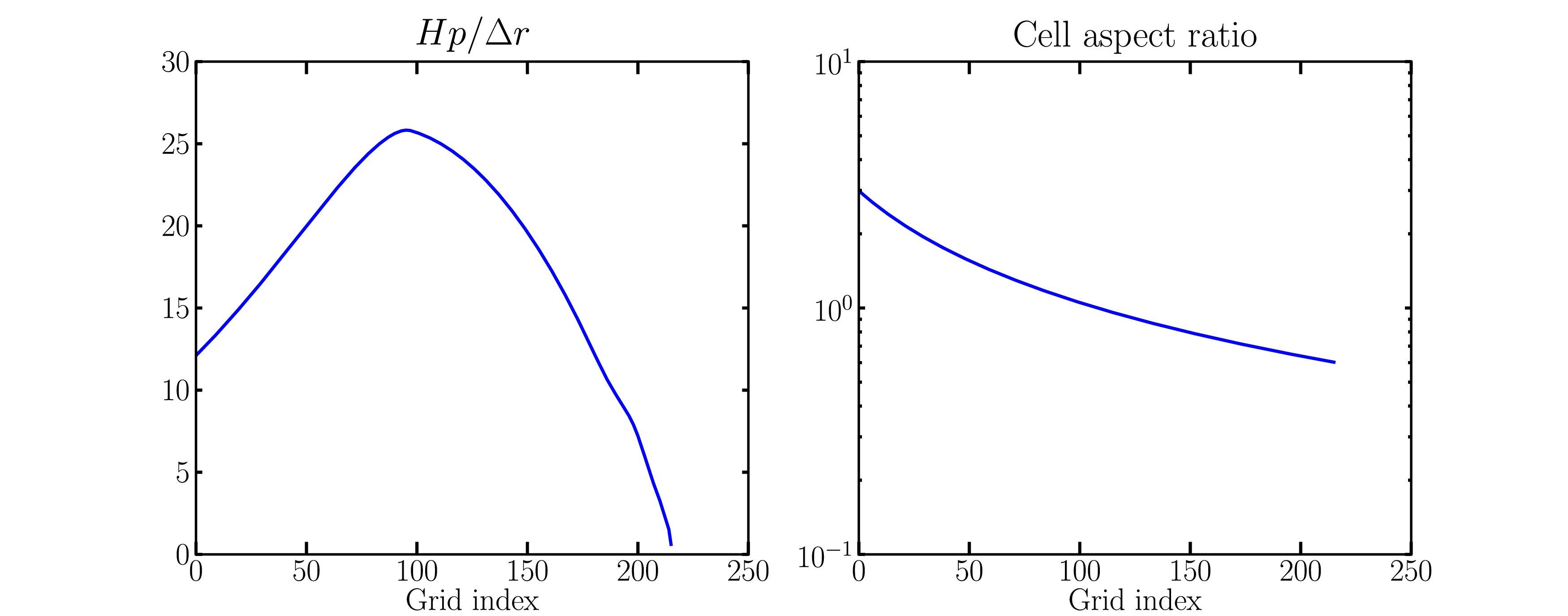} 
   \caption{Properties of the uniform grid for the giant model with $N_r=217$. Left panel: Pressure scale height to radial cell size ratio. Right panel: Cell aspect ratio when the radial mesh is extended in the tangential direction.}
   \label{fig:Cold_grid1_uniform}
\end{figure*}

\subsubsection{Initial model and grid design}

For this model, MLT predicts that about 50 \% of the stellar envelope is convective. The initial model for the multi-dimensional simulations is based on a 1D MLT model. Starting from a radiative model, as previously done for the A-type star, yields difficulties since a radiative structure departs strongly from the convective solution.
The MLT model might also depart from the \emph{real} convective state, but one can expect this difference to be less pronounced than with a fully radiative model. We model a significant fraction of the star (80 \%) in order to have a large radiative zone below the convective zone. Our numerical domain therefore extends radially from $0.2 R_\star $ to $R_\star$, where $R_\star$ is the radius of the star as defined by the location of the photosphere. This yields a pressure contrast of roughly six orders of magnitude ($\sim 14$ pressure scale heights) and a density contrast of almost five orders of magnitude throughout the whole domain. Throughout the convective zone the pressure contrast is $\sim 400$ ($\sim 6$ pressure scale heights) and the density contrast is $\sim 50$. Previous works that also consider an extended radial domain are the``star-in-a-box" simulations of \cite{2002AN....323..213F}, \cite{2003ASPC..293...45W}, and \cite{2006ApJ...638..336D}, where the full star is embedded in a Cartesian domain. \cite{2006ApJ...638..336D} consider a very simplified stellar model, with a very low-density stratification (a factor of $\sim10$). The two other works consider more realistic stellar models of super giant and AGB stars and obtain convection over multiple pressure scale heights. We also mention \cite{2009ApJ...702.1078B}, which presents a numerical simulation of the inner $\sim 50$ \% of a red giant star with the anelastic code ASH. Their density stratification in their convective zone (which occupies the whole numerical domain) is about two orders of magnitude.

Our initial model extends up to the photosphere. The photospheric region is characterised by the ionisation regions of HeII and HeI/H, with ``bumps" in the opacity profile. In addition, the surface region is characterised by the lowest values of the pressure scale height within the whole stellar structure. A non-uniform radial mesh would be necessary to solve the different scales present in the star. Our 1D stellar structure code uses such an adapted radial grid, which ensures a very good resolution of all features throughout the star. This is illustrated in the left panel of Fig. \ref{fig:Cold_grid1}, where we have sampled the initial 1D model to create a radial mesh with a suitable number of grid points (200 in this case). 

However, when considering multi-dimensional calculations, this grid presents a flaw: extension in the tangential direction yields intractable aspect ratio of the cells. This is illustrated in the right panel of Fig. \ref{fig:Cold_grid1}, which shows the radial profile of the cell aspect ratio when the former radial mesh is simply extended in the tangential direction by putting 256 cells over a tangential angular width of $\pi/2$. Inspection of Fig. \ref{fig:Cold_grid1} shows that a constant number of cells per scale height translates into a large variation of the cell aspect ratio. In the particularly demanding region near the photosphere the aspect ratio drops below $10^{-2}$, which is unacceptable for performing numerical simulations.

It is therefore obvious that to perform multi-dimensional simulations from the deep stellar interior to the surface, one has to make a compromise on the resolution when using a single grid. A solution to this problem would be to use a system of nested grids where the grids are refined in both the radial and tangential directions as moving near the surface of the star. Here, we choose to consider a computational grid with an uniform grid spacing in both $\Delta r$ and $\Delta \theta$ and apply a special treatment for the surface layers (see below). Due to spherical coordinates, the aspect ratio of the cells decreases with the radius. The radial step is chosen so that the cells in the middle of the domain are square. The right panel of Fig. \ref{fig:Cold_grid1_uniform} shows that our cells have an aspect ratio of 3 at the bottom of our domain and 0.6 at the top of the domain. As a result, the surface of the star is not resolved well, as shown by the number of grid points per pressure scale height (see left panel of Fig. \ref{fig:Cold_grid1_uniform}).

\subsubsection{Simulation setup}

Since we cannot accurately describe the surface of the star, we choose to apply a heating function that drives an isothermal region in the outer, badly resolved, region of the star. To do so, we include in our energy equation a heating source term on the right-hand side:

\begin{equation}
Q= - \rho c_v ( T - T_0 ) f(r) / \tau_\mathrm{heating},
\end{equation}

\noindent where $f(r)$ is a spatially varying function that is equal to 1 above a given radius $r_c$ and zero elsewhere, with a smooth transition in between, $\tau_\mathrm{heating}$ is the typical cooling time scale and $T_0$ is the forcing temperature. This method has been used for the ``star-in-box" simulation presented in \cite{2006ApJ...638..336D} (see also references therein). The parameter $T_0$ is chosen to be the temperature of the gas at $r_c$ in the initial model of the star. For the model presented in this section, $r_c=0.95 R_\star$, $T_0=32\,750$ K, and $\tau_\mathrm{heating} = 10^4$ s. For comparison, the value of the radiative time scale (as defined in Eq. \ref{parabolic_cfl}) decreases exponentially from $\sim 3.5\times10^5$ s to $\sim 3.5\times 10^3$ s in this region so our value of $\tau_\mathrm{heating}$ insures that the source term acts on a short enough time scale to keep the outer region in a roughly isothermal state.

For the value of $T_0$ used in our model, we do not have the HeI/H ionisation region near the surface. Also we prevent any convection motion in this region since the isothermal structure is stably stratified. There is essentially no radiative flux throughout this region, because the energy entering this region is evacuated by the source term instead. 

This treatment of the surface layers is very basic, and is only used as a preliminary solution to produce our first results. We are currently investigating other more physically based  methods to treat the surface layers.

\subsubsection{Results}

\begin{table*}[tb]
\caption{Numerical parameters for the cold giant run (same as Table \ref{table:Astar}).}
\label{table:cold}      % is used to refer this table in the text
\centering                          % used for centering table
\begin{tabular}{c c c c c c c c c}        % centered columns (4 columns)
\hline\hline                 % inserts double horizontal lines
Resolution & Domain & $\tau_\mathrm{KH}$ & $\Delta t_\mathrm{hydro}^\mathrm{CFL}$ & $\Delta t_\mathrm{rad}^\mathrm{CFL}$ & Final time & Time steps & Wall Time & Mem.\\
\hline                        % inserts single horizontal line
 $217\times 256$ & $[0.2,1.]\times[\pi/4,3\pi/4]$ & 2200 yr & 250 s & $3.2\times 10^{3}$ s & 4235 d & 16216 & 8 w &  2.2 Gb\\
\hline                         %inserts single line
\end{tabular}
\end{table*}

\begin{figure}[t] %  figure placement: here, top, bottom, or page
   \centering
   \vspace{0.5cm}
   \includegraphics[width=7cm]{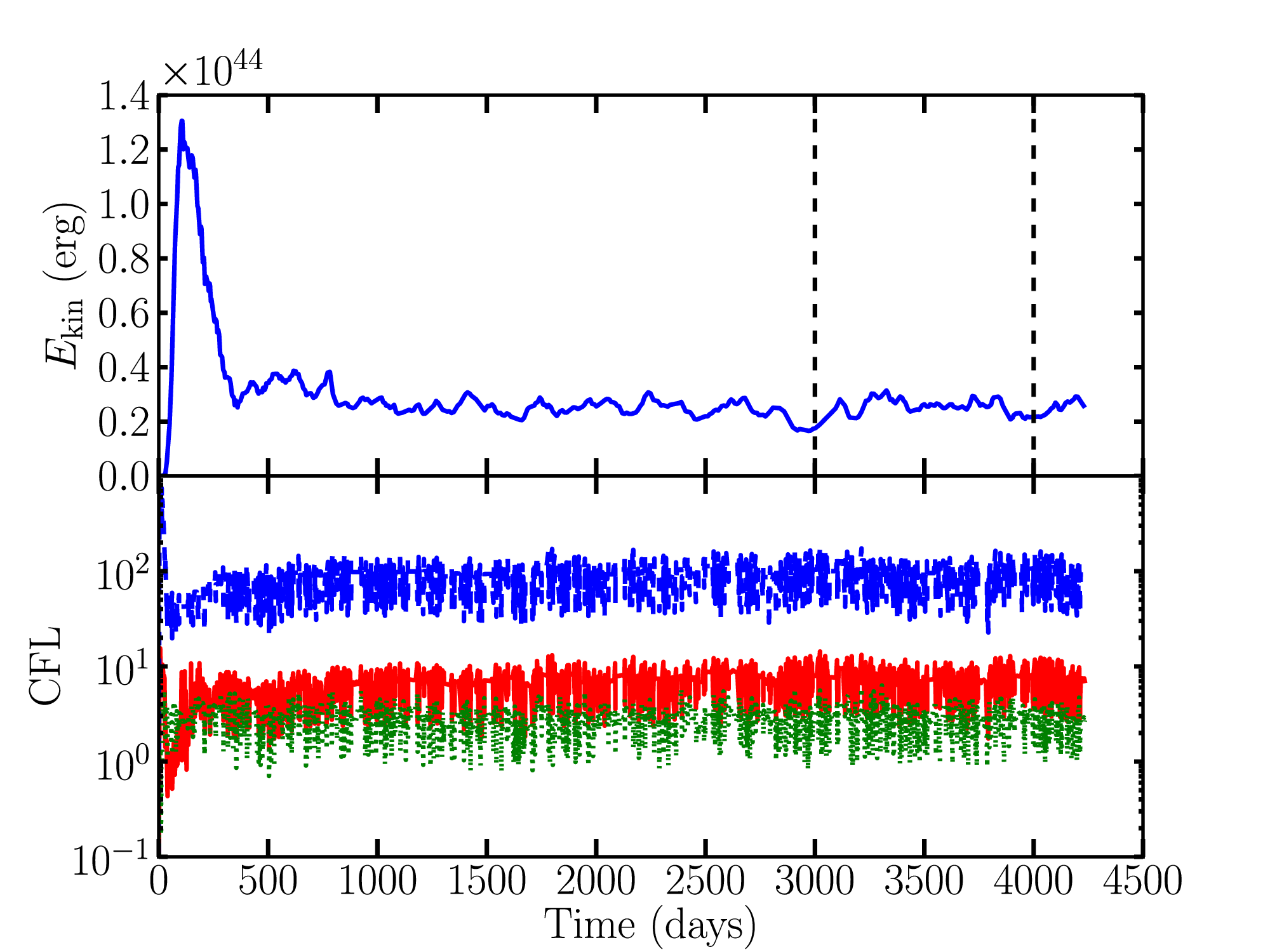} 
   \vspace{0.5cm}
   \caption{Evolution of the total kinetic energy (top panel) and CFL numbers for the giant star run (lower panel, solid red: radiative; dashed blue: hydro; dotted green: advection). The vertical dashed lines show the time interval on which the averaged fluxes shown in Fig. \ref{fig:cold_Fconv} were computed.}
   \label{fig:cold_Ekin_CFL}
\end{figure}

\begin{figure*}[t] %  figure placement: here, top, bottom, or page
   \centering
   \includegraphics[angle=90,width=0.8\linewidth,viewport=45 30 475 680, clip]{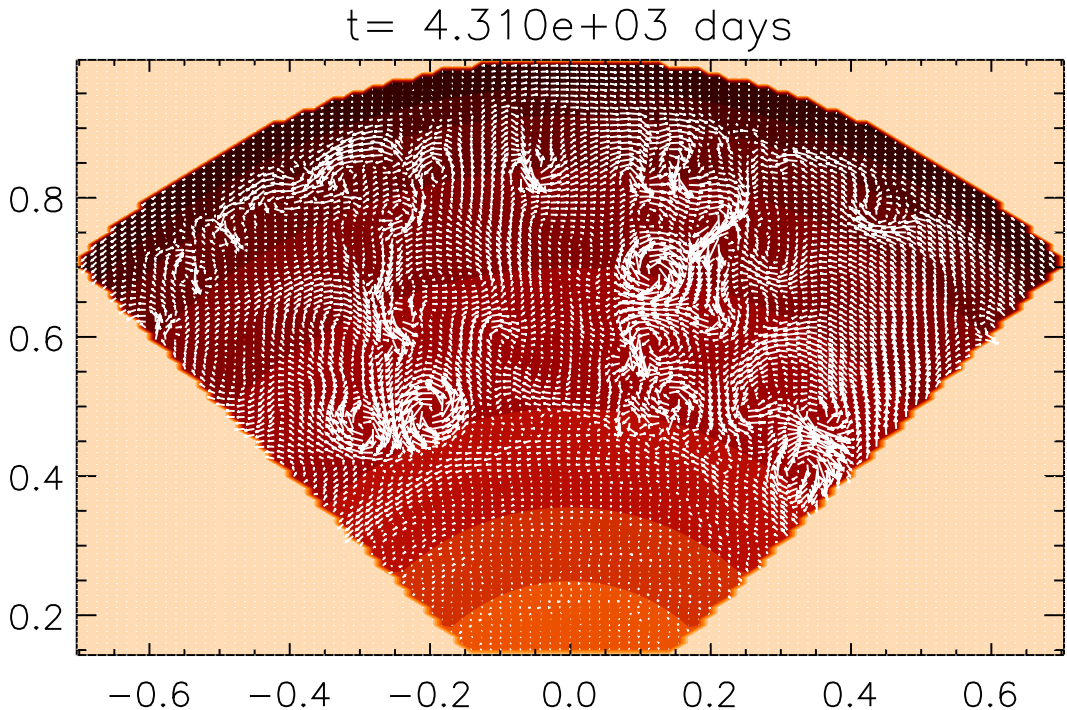} 
   \caption{Snapshot of the cold giant star simulation. The axis are Cartesian coordinates normalised by the stellar radius and with the origin at the star centre.}
   \label{fig:cold_SN}
\end{figure*}

We now present the results of the run (summarised in Table \ref{table:cold}). Figure \ref{fig:cold_Ekin_CFL} shows the evolution of the kinetic energy and CFL numbers. As in the previous section, we see that, during the initial phase of relaxation/development of convection, we reach high CFL numbers, with the hydro CFL number reaching a maximum value around $10^3$. As the convective instability develops, we see a quick and strong rise in the kinetic energy in the domain, followed by a slower decrease as the model relaxes toward equilibrium. After $500$ days, the kinetic energy tends toward a constant value, indicating that a quasi steady-state is reached. The time step remains roughly constant with a mean value of $\Delta t = 2.5\times 10^4 \pm 5\times 10^3$ s, corresponding to CFL$_\mathrm{hydro} = 100 \pm 20 $, and CFL$_\mathrm{rad}= 7.6 \pm 1.5$. The second value indicated corresponds to the standard mean deviation. Therefore in this kind of model, we see that the radiative time scale is much lower than the hydro time scale. Figure \ref{fig:cold_SN} shows a snapshot of the convective motions in the star.

\begin{figure}[t] %  figure placement: here, top, bottom, or page
   \centering
   \vspace{0.5cm}
   \includegraphics[width=7cm]{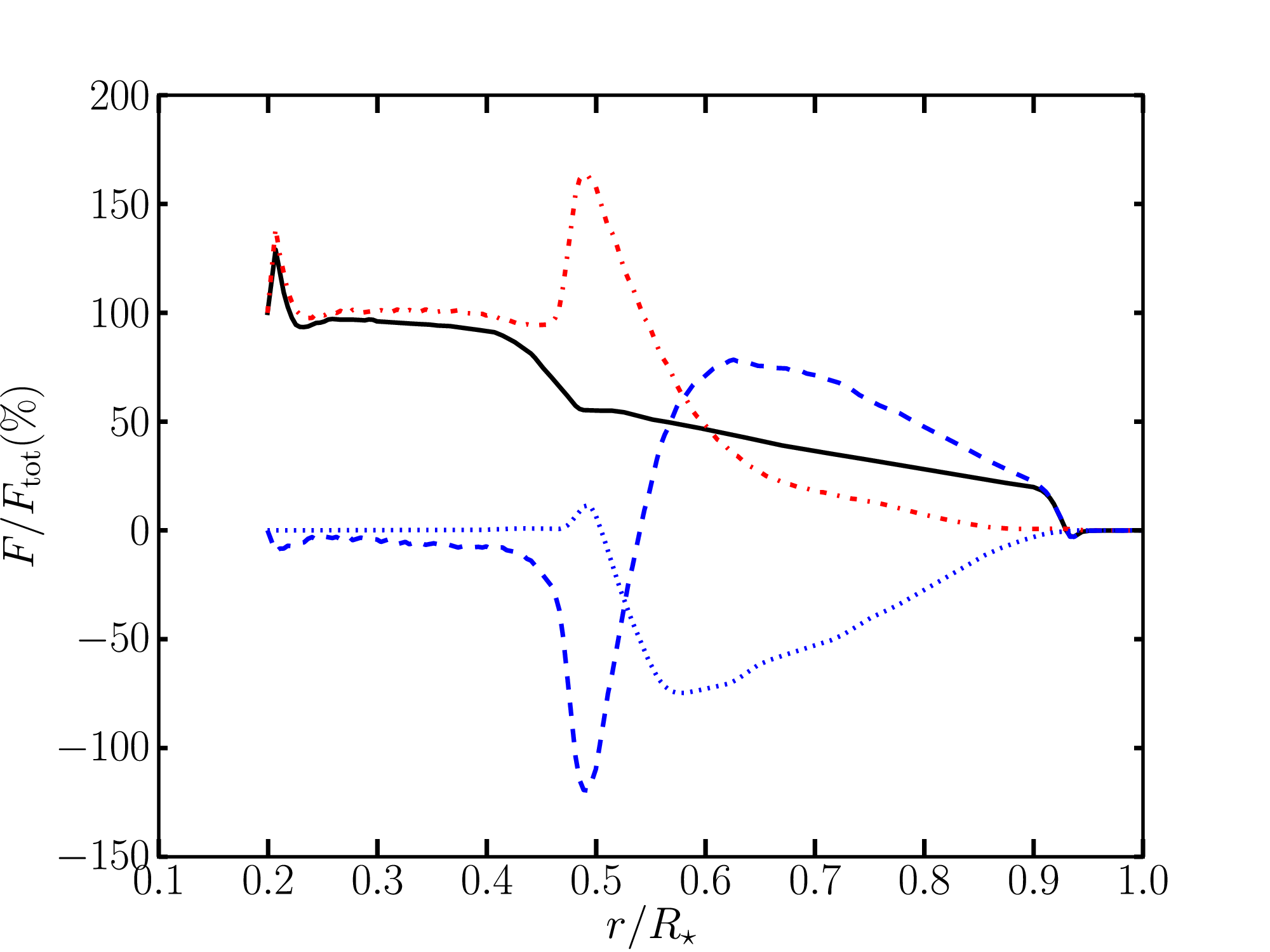} 
    \vspace{0.5cm}   
   \caption{Radial profiles of averaged fluxes for the giant star run (dashed-dotted line: radiative flux; dashed line: enthalpy flux; dotted line: kinetic energy flux; solid line: total flux). }
   \label{fig:cold_Fconv}
\end{figure}

\begin{figure}[t] %  figure placement: here, top, bottom, or page
   \centering
   \includegraphics[width=\linewidth]{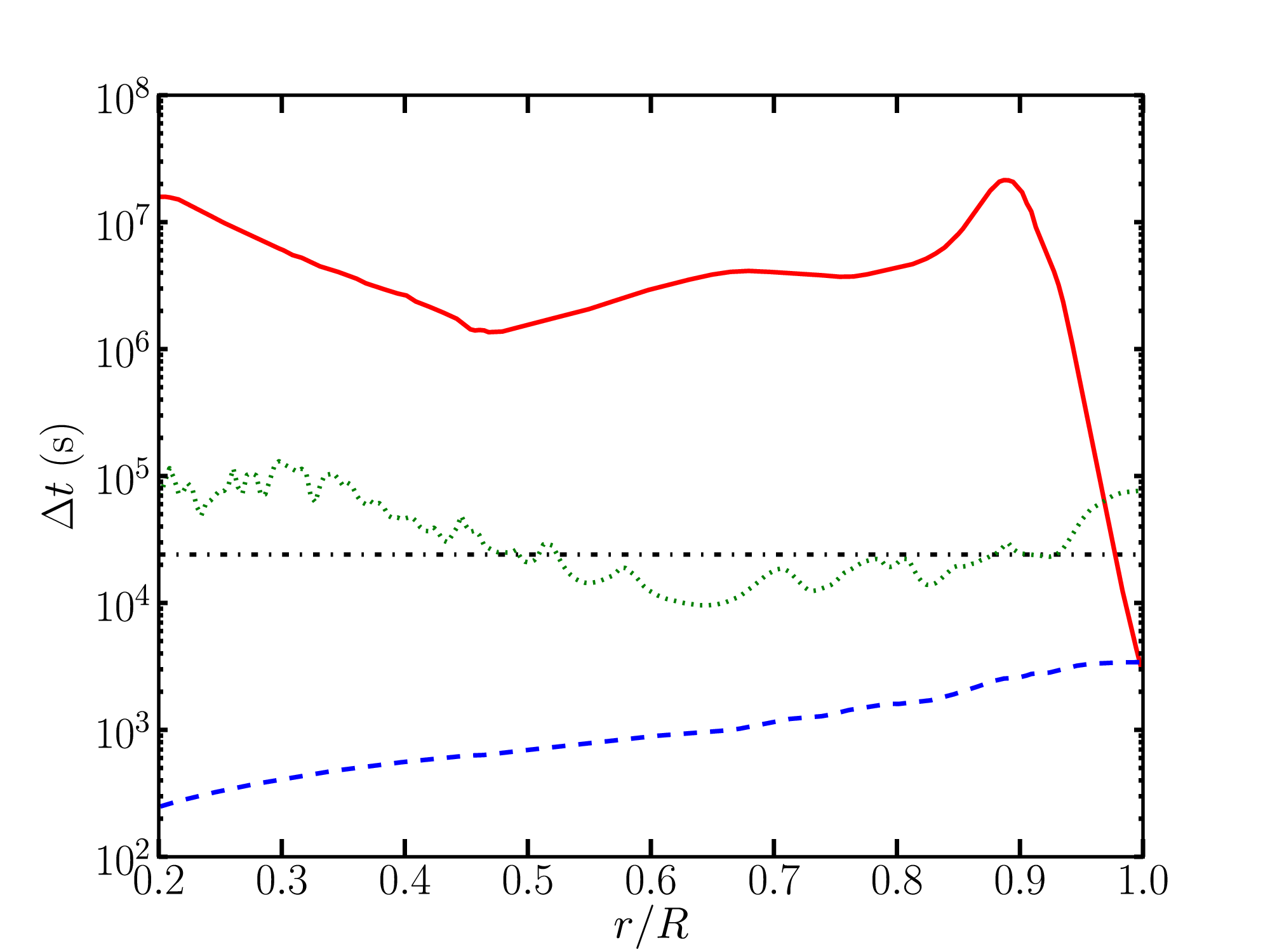} 
   \caption{Radial profiles of the CFL time steps for the giant star run (see text). From bottom to top: hydro (dashed blue), advection (dotted green), and radiation (solid red). The horizontal dash-dotted line indicates the time step of the snapshot under consideration.}
   \label{fig:cold_dt}
\end{figure}

We computed the averaged fluxes defined in Sect. \ref{Astar} between $t=3000$ days and $t=4000$ days (corresponding to 3452 time steps). The resulting radial profiles are shown in Fig. \ref{fig:cold_Fconv}. We see that thermal equilibrium has not been reached yet, since the total flux throughout the envelope is not constant. We see a significant upward enthalpy flux in the $\sim 40 \%$ outer radius of the star. Also, we see a very important downward kinetic energy flux in the convective zone. 
Around $r \sim 0.5 R_\star$, we observe strong bumps both in the radiative and in the enthalpy flux. This region corresponds to the interface between the convective zone and the radiative zone. When downdrafts cross the interface, the low efficiency of radiation prevent them from warming up,  and as a result, they are subject to a very strong braking by buoyancy forces, which prevent them to significantly overshoot in the radiative zone. This is why the interface region is strongly perturbed whereas the inner radiative zone remains unperturbed. Finally, we checked the range of Mach numbers in the domain. The average Mach number of the flow extends from $\sim 10^{-3}$ at the bottom of the envelope to $\sim 0.15$ at the top of the convective zone. However, Mach numbers in the range $ 0.30 - 0.40$ can be reached locally, in the downdrafts. We stress that such a range of Mach number cannot be addressed by anelastic codes, which are restricted to low Mach number flows (M$_s \lesssim 10^{-2} - 10^{-1}$).

%We explain these strong downwards fluxes as being related to the relaxation process. One should bear in mind that the current simulation ran for 1600 days of stellar time, whereas the thermal time scale of the modelled envelope is equal to 2200 years. 

Figure \ref{fig:cold_dt} shows the radial profiles of the different CFL time steps (as in Fig. \ref{fig:Astar_dt}). As already mentioned above, we see that the radiative CFL time step is much longer than the hydro time step; however, the strong decrease in the radiative time scale in the isothermal surface layer comes from the decrease in density in this region. We can expect that, if the surface regions were better resolved, the radiative time scale would drop to a lower value than the hydro time step, and we would recover locally a situation similar to the A-type star model. Here again the shortest hydro CFL time step is found at the bottom of the envelope, therefore we reach a higher hydro CFL number than for the A-type star model because our model extends deeper in the star. Finally, as discussed in the previous section, we see that the actual time step is roughly equal to the lowest value of the advective time scale, which is located in the convective zone. In this case, since the envelope extends deeper, our implicit scheme allows for reaching high values of the hydro CFL number by removing the limitation coming from the deeper layers of the star.

The results presented in this section show that it is meaningful to adopt a fully implicit approach for computing models of deep convective zones. Indeed, we show here an example of a situation where substantial CFL numbers could be reached by overcoming the time step limitation imposed by deep layers of the stellar envelope. This is possible since the flow velocity is low in those regions. In the surface convective region, however, we are still prone to the time step limitation imposed by the flow movement across grid cells. 

Finally, the computation of such deep stellar envelopes also poses the problem of how to achieve the thermal relaxation of the convective envelope on an acceptable time scale. Indeed, the thermal time scale of our numerical domain is

\begin{equation}
\tau_\mathrm{KH} = \frac{E_\mathrm{int}}{L} \sim 2200 \mathrm{\ years},
\end{equation}

\noindent whereas the simulation only ran for 4235 days $\sim 11.6$ years of stellar time. This is, however, an unavoidable difficulty that is common to all numerical approaches. 

\section{Conclusion}
\label{conclusion}

We have presented a new numerical code devoted to stellar interiors. The code solves the full set of the compressible Euler equations and radiative diffusion with a spatially and temporally accurate implicit scheme. 

As a first step, we computed stellar convection following two approaches: a local approach for a relatively hot star (see Sect. \ref{Astar}) and a global approach for a cold giant star (see Sect. \ref{giant}). In the first case, we were able to simulate the convective motions in the outer layers of a A-type star. For this type of model, we achieve a mean CFL number of 10 for hydro and of 400 for radiation. The rather low value of the hydro CFL number suggests that, in this particular case, a mixed approach (explicit hydro + implicit radiation) would result in a substantial speeding up of the calculation compared to a fully explicit or implicit approach. In the second case, we modelled the turbulent convective motions that transport energy in the bulk of a cold giant star. Our domain includes both the outer convective region and the bottom, stable, radiative region. For this model, we could reach a mean CFL number of 100 for hydro and of 7 for radiation. Such values of the hydro CFL numbers are encouraging and suggest that we might gain computational time thanks to our implicit approach. %However, this remains to be proven by performing benchmarking against an explicit version of our code.

The aim of this paper is to show the {\it feasibility} of using a fully implicit method for the accurate computation of time-dependent flows. We stress that our goal is, however, not to demonstrate that the code is successful in the sense of efficient production. Implicit methods are computationally much more expensive than explicit methods because they need to solve a set of non-linear equations. The efficiency of an implicit code is essentially set by the efficiency of the method used to solve this difficult task. The general framework for solving the non-linear system is the Newton-Raphson method. For tests and development purposes, we chose here a very simple approach where the Jacobian matrix is computed at each iteration, and where the linear system is solved \emph{exactly}  inside the Newton-Raphson loop. This results in a robust, but expensive non linear solver. A more efficient class of non linear solvers include \emph{inexact Newton-Krylov methods}, which are Newton-Raphson methods where an iterative method is used to solve the linear system, see e.g. see \citealt{Knoll2004357} for a review. These methods are Jacobian free methods and gain computational time by using an iterative method to \emph{approximately} solve the linear system during the Newton-Raphson loop (hence the name ``inexact"). These methods are very popular in computational physics and we plan to implement and test them in a future version of the code. 
Once the most efficient non-linear solvers currently available are implemented, we will be able to benchmark our code against explicit ones.

Concerning our numerical method, we found that the maximum time step that we can use without being faced with convergence issues in the Newton-Raphson procedure is closely related to the fluid velocity in the numerical domain. We introduced the \emph{advective time scale}

\begin{equation}
\label{eq:tadv}
t_\mathrm{adv} = \frac{\Delta x}{|u|},
\end{equation}

\noindent and showed that for the computation of a time-dependent flow the time step cannot grow much above the minimum value of this time step. The definition of Eq. \ref{eq:tadv} should be adapted in the multi-dimensional case. In other words, our numerical method cannot handle fluid movements across several cells, which is intuitively and physically understandable. We stress that such a criteria is also suggested by accuracy concerns. This limitation on the time step stems from the fact that in order to advance the solution over one time step, one has to solve a system of non-linear equations with a Newton-Raphson procedure. For time step larger than the advective time scale, the fluid moves across several cells and the solution changes significantly between two time steps. This change is essentially the limiting factor for increasing the time step, since the Newton-Raphson procedure encounters problems when relative changes between two time steps are too large. We stress that one could improve this by finding a better initial guess for the Newton-Raphson procedure, instead of simply taking the solution of the last time step. In 1D implicit simulations, a time extrapolation from preceding time steps has been used successfully to produce an initial guess for the Newton procedure. However, in multi-dimensional flows, this procedure does not necessarily gives a satisfactory guess of the solution. Indeed our experience did not show any improvement by using such an extrapolated guess.

Our first developments show that using a fully implicit code to describe stellar interiors is realistic, providing a powerful tool for studying a full stellar structure characterised by low-to-moderate Mach numbers. Our test performed on the envelope of a giant star indeed highlights the possibility to extend the spatial domain to a large fraction of the star (80\% in radius), where Mach numbers cover values from less than 1\% to a few tens of percent. More generally, this implicit tool will be efficient for computing relaxation toward steady state solutions and very low Mach number flows,  and to avoid the CFL constraint due to a particular location of the numerical domain. In the later case, it is possible to have a moderate Mach number flow in another region of the domain (this region would then be responsible for the limitation of the time step), as illustrated by the red giant case mentioned above.

Our first developments also stress the numerical difficulty of modelling a full star with a grid-based code. Since the typical length scales (e.g. pressure scale height) vary significantly between the deep interior of the star and the surface, it is not possible to resolve both the interior and the surface layers on a single grid. We plan to overcome this difficulty by implementing a system of nested grids refined in both the radial and tangential directions. Among other future developments of the code, we plan to extend it to 3D and to implement iterative solvers, rather than a direct method, involving parallelisation. Among our first astrophysical applications, we quote (i) the very early stages of evolution of low-mass stars, bridging the gap between the hydrodynamical phase of star formation and the quasi-static pre-main sequence evolutionary phase, (ii) the study of convection in pulsating stars, and (iii) the effect of fast rotation on convection (in low-mass stars,  massive stars). But clearly, a wide range of astrophysical problems can be studied with such a tool. Based on our experience, we hope to achieve these future developments and the first astrophysical applications within a reasonable period. We also hope that our efforts demonstrate the feasibility of applying an implicit multi-dimensional hydrodynamical code to stellar interiors and will motivate the computational astrophysics community to develop such a tool.

\begin{acknowledgement}
MV acknowledges support  from a Newton International Fellowship during part of this work. The numerical simulations were achieved thanks to the resources of the PSMN (``P\^ole Scientifique de Mod\'elisation Num\'erique") at the ENS de Lyon. The authors are indebted to C\'edric Mulet-Marquis and Emmanuel L\'ev\^eque for their significant contribution to an earlier version of the code. The authors are particularly thankful to Bernd Freytag and Ewald Mueller for many valuable and useful discussions which considerably helped the elaboration of this work. We also thank Sacha Brun for discussion and advice. Part of this work was funded by the European Research Council under the European CommunityÕs 7th Framework Programme (FP7/2007-2013 Grant Agreement no. 247060) and by the French ``Programme National de Physique Stellaire" (PNPS). Finally, the authors thank the anonymous referee for his/her useful comments that helped to improve the paper.
\end{acknowledgement}

\bibliographystyle{aa}
\bibliography{biblio}

%\newpage
\appendix

\section{Tests cases}
\label{appendix:tests}

\subsection{Linear test cases}

\begin{table*} 
\caption{Linear advection, $\Delta t = 10^{-1}$.} 
\label{table:adv1}	% is used to refer this table in the text
\centering 
\begin{tabular}{c | c c | c c | c | c} 
\hline
N & $L_1$-error & Rate & $L_\infty$-error & Rate & CFL & Wall Time (s) \\	% table heading
\hline 
   50 & 1.380e-02  &   -  & 1.152e-02 &     -  & 7.80e-01 & 0.9\\
  100 & 3.457e-03  & 2.00 & 2.587e-03 & 2.16 & 1.58e+00 & 1.8\\
  200 & 3.138e-03  & 0.14 & 7.491e-04  &1.79 & 3.17e+00 & 3.5\\
  400 & 3.252e-03 & -0.05 & 8.115e-04 & -0.12 & 6.35e+00 & 7.7\\
  800 & 3.308e-03 & -0.02 & 8.269e-04 & -0.03 & 1.27e+01 & 27.5\\
\end{tabular} 
\end{table*}

\begin{table*} 
\caption{Linear advection, $\Delta t = 10^{-2}$.} 
\label{table:adv2}	% is used to refer this table in the text
\centering 
\begin{tabular}{c | c c | c c | c | c} 
\hline
N & $L_1$-error & Rate & $L_\infty$-error & Rate & CFL & Wall Time (s) \\	% table heading
\hline 
   50 & 1.890e-02 &    -  & 1.292e-02 &  -  & 7.80e-02 & 3.8\\
  100 & 4.481e-03  & 2.08 & 4.882e-03  & 1.40 & 1.58e-01 & 7.5\\
  200 & 1.027e-03  & 2.13 & 1.830e-03  & 1.42 & 3.17e-01 & 15.3\\
  400 & 2.094e-04  & 2.29 & 6.732e-04  & 1.44 & 6.35e-01 & 34.5\\
  800 & 3.996e-05  & 2.39 & 1.936e-04  & 1.80 & 1.27e+00 & 134.2\\
  \end{tabular} 
\end{table*}

\begin{table*} 
\caption{Linear advection, $\Delta t = 10^{-3}$.} 
\label{table:adv3}	% is used to refer this table in the text
\centering 
\begin{tabular}{c | c c | c c | c | c} 
\hline
N & $L_1$-error & Rate & $L_\infty$-error & Rate & CFL & Wall Time (s) \\	% table heading
\hline 
   50 & 1.895e-02 &    -  & 1.293e-02   &  -  & 7.80e-03 & 35.2\\
  100 & 4.530e-03 & 2.06 & 4.892e-03 & 1.40 & 1.58e-02 & 68.5\\
  200 & 1.074e-03 & 2.08 & 1.845e-03 & 1.41 & 3.17e-02 & 134.4\\
  400 & 2.569e-04 & 2.06 & 6.918e-04 & 1.42 & 6.35e-02 & 273.9\\
  800 & 6.102e-05 & 2.07 & 2.569e-04 & 1.43 & 1.27e-01 & 1016.7\\
\end{tabular} 
\end{table*}

\begin{table*} 
\caption{Linear diffusion, $\Delta t = 10^{-2}$.} 
\label{table:diff1}	% is used to refer this table in the text
\centering 
\begin{tabular}{c | c c | c c | c | c} 
\hline
N & $L_1$-error & Rate & $L_\infty$-error & Rate & CFL & Wall Time (s) \\	% table heading
\hline 
   50 & 2.072e-03  &   -  & 1.824e-03 &   -  & 1.50e+00 & 0.4\\
  100 & 5.237e-04  & 1.98 & 3.659e-04 & 2.32 & 6.13e+00 & 0.6\\
  200 & 8.663e-04 & -0.73 & 8.220e-04 & -1.17 & 2.48e+01 & 1.0\\
  400 & 9.727e-04 & -0.17 & 9.487e-04 & -0.21 & 9.95e+01 & 2.0\\
  800 & 9.999e-04 & -0.04 & 9.802e-04 & -0.05 & 3.99e+02 & 6.6\\
   \end{tabular} 
\end{table*}

\begin{table*} 
\caption{Linear diffusion, $\Delta t = 10^{-3}$.} 
\label{table:diff2}	% is used to refer this table in the text
\centering 
\begin{tabular}{c | c c | c c | c | c} 
\hline
N & $L_1$-error & Rate & $L_\infty$-error & Rate & CFL & Wall Time (s) \\	% table heading
\hline 
   50 & 2.825e-03  &  -  & 2.812e-03  &  - & 1.50e-01 & 2.6\\
  100 & 6.874e-04  & 2.04 & 6.743e-04 &  2.06 & 6.13e-01 & 4.9\\
  200 & 1.629e-04  & 2.08 & 1.586e-04  & 2.09 & 2.48e+00 & 9.2\\
  400 & 3.419e-05  & 2.25 & 3.158e-05  & 2.33 & 9.95e+00 & 18.3\\
  800 & 4.973e-06  & 2.78 & 3.406e-06  & 3.21 & 3.99e+01 & 64.4\\
\end{tabular} 
\end{table*}

\begin{table*} 
\caption{Linear diffusion, $\Delta t = 10^{-4}$.} 
\label{table:diff3}	% is used to refer this table in the text
\centering 
\begin{tabular}{c | c c | c c | c | c} 
\hline
N & $L_1$-error & Rate & $L_\infty$-error & Rate & CFL & Wall Time (s) \\	% table heading
\hline 
   50 & 2.833e-03  &  -  & 2.823e-03 &  -  & 1.50e-02 & 24.8\\
  100 & 6.963e-04  & 2.02 & 6.847e-04 &  2.04 & 6.13e-02 & 47.9\\
  200 & 1.717e-04  & 2.02 & 1.690e-04 & 2.02 & 2.48e-01 & 91.4\\
  400 & 4.265e-05  & 2.01 & 4.193e-05 & 2.01 & 9.95e-01 & 183.7\\
  800 & 1.059e-05  & 2.01 & 1.038e-05 & 2.01 & 3.99e+00 & 640.5\\
\end{tabular} 
\end{table*}

\subsubsection{Advection equation}

In this section we test our numerical scheme with the 1D scalar advection equation:

\begin{equation}
\label{eq:advection}
\frac{\partial q}{\partial t} + a \frac{\partial q}{\partial x} = 0.
\end{equation}

\noindent We use a smooth profile as an initial condition:

\begin{equation}
f(x) = \sin(x),\ x \in [0, 2\pi].
\end{equation}

\noindent The advection velocity $a$ is set to 1. We advect the sin wave for one unit of time and the resulting numerical solution $q_{i+1/2}$ is compared with the exact solution $q^0_{i+1/2}=f(x_{i+1/2}-1)$ by computing the $L_1$ and $L_\infty$ errors:

\begin{eqnarray}
|| q - q^0||_1 &=& \Delta x \sum_i | q_{i+1/2} - q^0_{i+1.2} | \\
|| q - q^0||_\infty &=& \max_i | q_{i+1/2} - q^0_{i+1.2} |.
\end{eqnarray}

For a given time step and mesh size, the CFL number is defined as $a \Delta t / \Delta x$. We use an uniform discretisation in space and perform a sequence of runs with different resolutions ranging from 50 to 800 grid points by successive increase by a factor of 2. This is done for three values of the time step in order to investigate a wide range of CFL: $\Delta t=$ $10^{-1}$, $10^{-2}$, $10^{-3}$. We use the Van Leer limiter and the Crank-Nicholson method, which are both second-order methods.

Such a grid of simulations allows to study the convergence properties of our code. The actual order $r$ of the scheme between two consecutive resolutions is computed as

\begin{equation}
r = \log \frac{|| q - q^0||^{(1)}}{|| q - q^0||^{(2)}} / \log 2,
\end{equation}

\noindent where (2) corresponds to a resolution twice higher than (1). The results are shown in Table \ref{table:adv1}, \ref{table:adv2}, and \ref{table:adv3}.

Before discussing the results, it is useful to derive a result that will help for their interpretation. Going back to Eq. (\ref{eq:advection}), the simplest implicit scheme that can be designed combines the first order backward Euler method with the first-order upwind method: 

\begin{equation}
\label{eq:firstorder}
q^{n+1}_i = q^n_i - \frac{a \Delta t}{\Delta x} \Big( q^{n+1}_i - q^{n+1}_{i-1} \Big ).
\end{equation}

For this scheme, it is straightforward to derive the so-called ``modified equation" (see e.g. \citealt{Hirsch:1990book2}):

\begin{equation}
\label{eq:advection_equ}
\frac{\partial q}{\partial t} + a \frac{\partial q}{\partial x} = \frac{a}{2} (\Delta x + \Delta t) \frac{\partial^2 q}{\partial x^2} + \mathrm{H.O.T.},
\end{equation}

\noindent which shows that formula \ref{eq:firstorder} solves an advection-diffusion problem since the leading term of the truncature error is physically equivalent to a viscous term with a numerical viscosity coefficient $D = \frac{a}{2} (\Delta x + \Delta t)$. Usually, with explicit schemes in mind, it is supposed that the ratio $\Delta x/\Delta t$ is kept constant (i.e. constant CFL number) so that $D \propto \Delta x$ or $D \propto \Delta t$, which is equivalent to the statement that scheme \ref{eq:firstorder} is first order in time and in space. On the other hand, when an implicit scheme is used, we cannot make such an assumption for the ratio $\Delta x/\Delta t$, as the CFL number is in principle arbitrary. In our discussion we use the expression of this numerical viscosity coefficient to interpret our results.

It appears that for CFL values above 1 (e.g. $\Delta t = 10^{-1}$, see Tab. \ref{table:adv1}), the numerical error remains roughly constant when the resolution is increased. This suggests that, in this case, the time truncation error dominates the spatial one. With the expression of the numerical viscosity derived above, we are in a regime where $D = \frac{a}{2} \Delta t = cte$, so when the resolution is increased the numerical dissipation does not decrease. Another way of interpreting this result is that the solution is advected over more than one cell width during one time step so that a strong numerical damping results. On the other hand, for low values of the CFL number (e.g. $\Delta t = 10^{-3}$, see Table \ref{table:adv3}), the spatial truncation error dominates, and we recover the second-order convergence of our spatial scheme in the $L_1$ norm. We do not see an accumulation effect of the round-off errors for very small time step. Again using the numerical dissipation derived above, one has for a first-order scheme $D = \frac{a}{2} \Delta x$. We expect that for our second-order scheme, a similar calculation would lead to $D \propto \Delta x^2$, in agreement with the observed tendency. A lower rate of convergence measured in the $L_\infty$ is characteristic of the action of the limiter on the smooth extrema of the solution. This is a well known drawback of TVD schemes: their order decreases also at smooth extrema of the solution. Finally for $\Delta t = 10^{-2}$, we are in the regime where $\Delta x \sim \Delta t$, so when the mesh size is decreased, the numerical dissipation decreases in a non trivial fashion. For instance, considering the first-order numerical dissipation derived above, decreasing the mesh  size by a factor of two would result in a decrease in the numerical dissipation by a factor of $\frac{1+\lambda}{1+\lambda/2}$, where $\lambda= \Delta x /\Delta t$. As a result the scheme does not behave as having a specific order in this case.

To summarise these results, we stress that the error in the numerical solution of partial derivative equations depends on both the spatial and temporal discretisation. When the temporal (resp. the spatial) truncation errors dominates, i.e. for very large (resp. very small) CFL numbers, decreasing the spatial (resp. temporal) step does not lead to more accurate results. It is therefore essential to choose both the mesh size and the time step in order to optimise the accuracy and the computational cost. Finally, we stress that the advection equation is not a suitable problem for implicit methods, since the stability criterion for an explicit scheme, namely $a \Delta t < \Delta x$, is also an accuracy criterion. The optimum accuracy/computational cost is therefore obtained for CFL number of order unity. Any advection computation at high CFL numbers suffers from serious numerical damping and lack of accuracy.

\subsubsection{Diffusion equation}

In this section, we test our diffusion scheme with the simple linear diffusion equation:

\begin{equation}
\label{eq:diff}
\frac{\partial T}{\partial t} =  \chi \frac{\partial^2 T}{\partial x^2}.
\end{equation}

\noindent For an initial Dirac distribution of amplitude $Q$ at x=0, the analytic solution of Eq. (\ref{eq:diff}) reads \citep[see e.g.][]{1984oup..book.....M} as

\begin{equation}
\label{eq:sol}
T(x,t) = \frac{Q}{\sqrt{4 \pi \chi t}} \exp \big ( -\frac{x^2}{4 \chi t} \big ).
\end{equation}

\noindent The problem is solved numerically on a uniform grid spanning the domain $[-2,2]$. For a given mesh size and time step, the CFL number is defined as $\chi \Delta t/\Delta x^2$. We use our second-order spatial scheme for diffusion, together with the second-order Crank-Nicholson scheme for time stepping.

For simplicity, we consider here $\chi = Q = 1$. Since it is not possible to represent a Dirac distribution exactly on a discrete mesh, we adopt as an initial condition the solution of Eq. (\ref{eq:sol}) at t=0.025. The solution is evolved until t=1, and we compare the numerical solution with the analytic solution, as done in the previous section for the linear advection problem. Again we perform a sequence of grid resolutions ranging from 50 to 800 grid points. This is done for three different time steps $\Delta t=$ $10^{-2}$, $10^{-3}$, and $10^{-4}$, yielding a wide range of CFL numbers. The results are shown in Tables \ref{table:diff1}, \ref{table:diff2} and \ref{table:diff3}.

As in the previous section, when the time step is large ($\Delta t = 10^{-2}$, see Table \ref{table:diff1}), the temporal truncation error dominates, and the numerical error does not decrease when the resolution is increased. On the other hand, for small time steps (e.g. $\Delta t = 10^{-4}$, see Table \ref{table:diff1}), the spatial truncation error dominates, and we recover a second-order convergence. In this case, both the $L_1$ and the $L_\infty$ show second-order behaviour, since there is no limiting process in the construction of the interface fluxes. Again in the transition regime  ($\Delta t = 10^{-3}$, see Table \ref{table:diff1}), we do not observe a well-defined order. Such behaviour is likely for a similar reason to the advection case.

Here, an implicit scheme is more efficient than an explicit scheme since, for a given resolution, for instance N=800, the run with $\Delta t=10^{-2}$ has a solution that is more accurate (by more than a factor of 10 in both norms) than the run with $\Delta t=10^{-3}$ (closer to the explicit time step), and has a computation time that is ten times shorter. In this case, the optimum ratio between accuracy and computational cost is obtained for CFL numbers larger than 1, because here the CFL condition is not an accuracy criterion.

\subsection{Non-linear test cases}

\subsubsection{Sod test}

A typical test for non-linear hydro is the 1D Sod shock-tube test. This is a Riemann problem with the following initial right and left states: $\rho_L=1$, $p_L=1$, $\rho_R=0.125$, $p_R=0.1$; the gas is initially at rest. The solution of the Sod problem is computed until $t=0.25$ on the domain $[-0.5,0.5]$, where the interface between the initial states is located at $x=0$. The number of grid points is 400. For simplicity and since the goal is to show how our code deals with shocks, we set the time step to keep the hydrodynamic CFL equal to 1. 

\begin{figure*}[p] %  figure placement: here, top, bottom, or page
  \centering
  \parbox{0.45\linewidth}{\includegraphics[width=0.95\linewidth,viewport=20 20 550 650,clip]{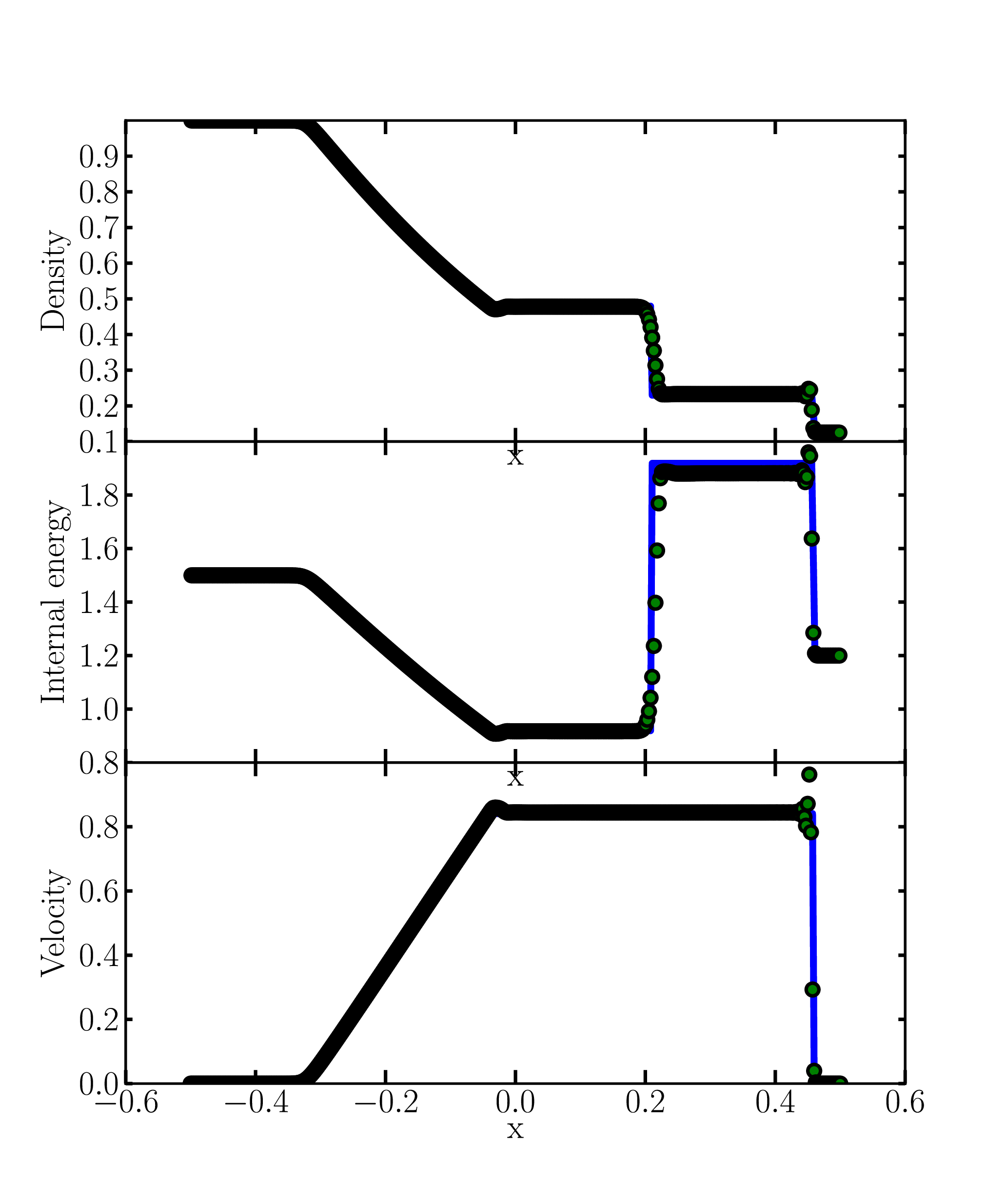}}
  \parbox{0.45\linewidth}{\includegraphics[width=0.95\linewidth,viewport=20 20 550 650,clip]{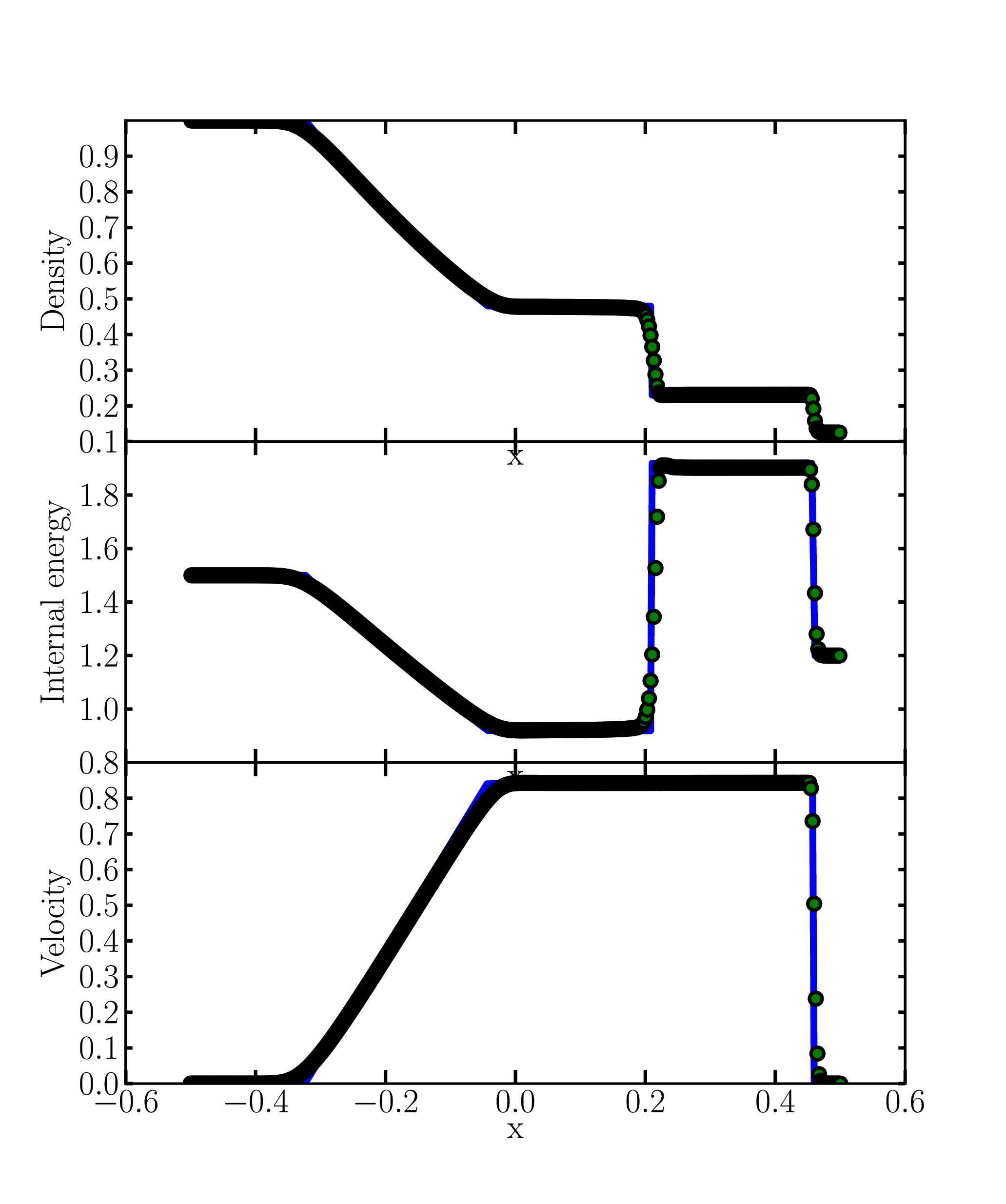}}
   \caption{Results of the Sod test at $t=0.25$ without artificial viscosity (left panel) and with $C=0.5$ (right panel). The dots are the numerical solution, the solid line is the analytical solution. The time step corresponds to CFL$_\mathrm{hydro}=1$.}
   \label{fig:sod}
\end{figure*}

The left panel of Fig. \ref{fig:sod} shows the result of the computation. One can easily identify the shock front, the contact discontinuity, and the rarefaction wave. The location and amplitude of these features are in excellent agreement with the analytical solution. However, one can see that spurious numerical effects affect the solution. Oscillations are produced at the location of the shock, despite the use of a limiter. An over/under shoot is also visible at the tail of the rarefaction wave. Solving the total energy equation and/or limiting the primitive variables (i.e. pressure) could improve these results. Otherwise, a possible cure is to introduce an artificial viscosity of the form:

\begin{equation}
\nu = C \Delta x c_s^2,
\end{equation}

\noindent where $c_s$ is the sound speed, $\Delta x$ the mesh size, and $C$ a dimensionless parameter that has to be fine-tuned to obtain acceptable results. The right panel of Fig. \ref{fig:sod} shows the result of the computation when $C=0.5$. We stress that we do not encounter shocks in our current calculations of stellar models.

\begin{figure*}[p] %  figure placement: here, top, bottom, or page
  \centering
  \includegraphics[width=0.8\linewidth,clip]{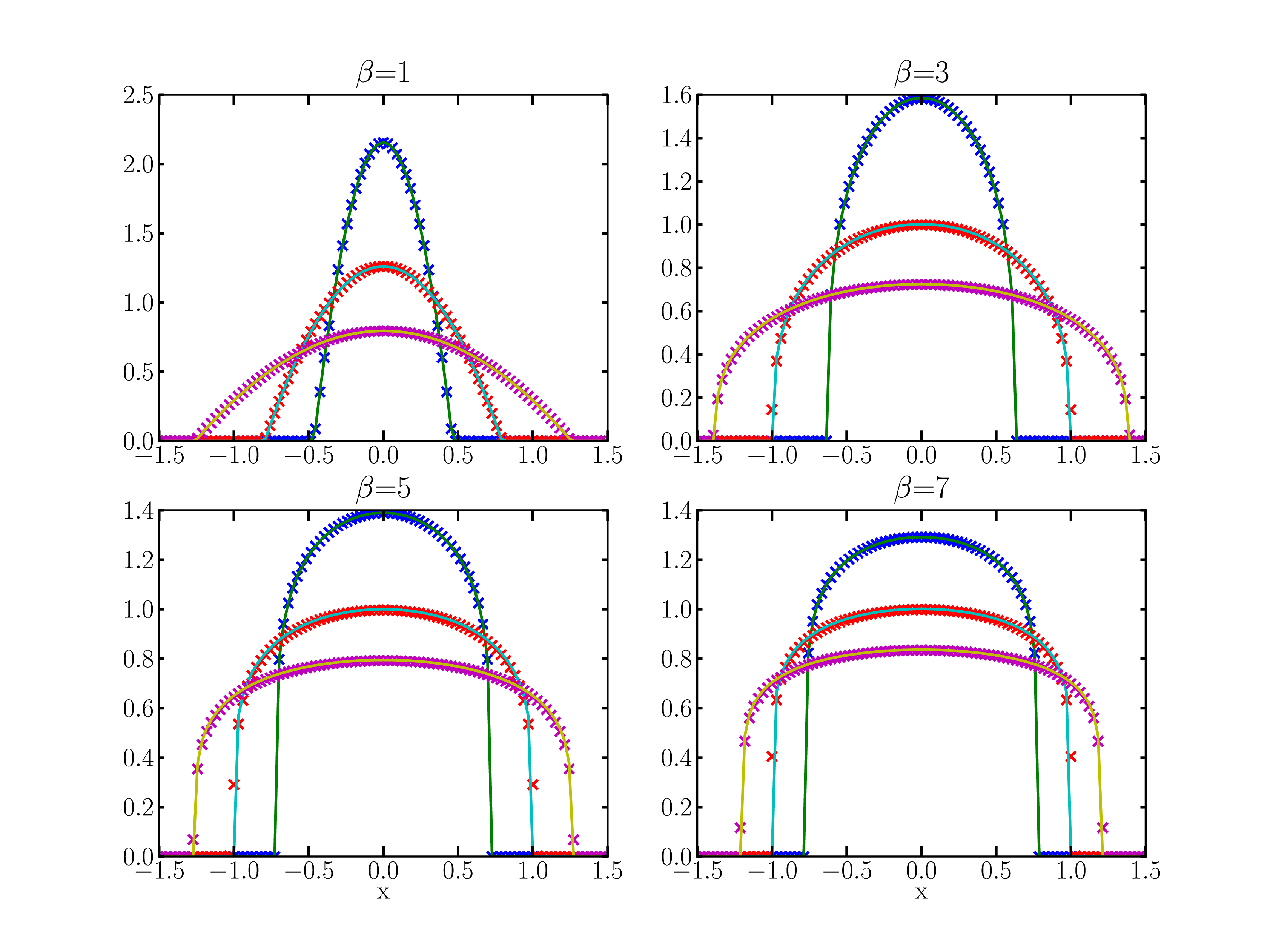}
   \caption{Results of the Barenblatt problem for $\beta$ = 1, 3, 5, 7. The solution is shown at $\bar{t}=$ 0.1, 1,  2 for $\beta=1$ and at $\bar{t}=$ 0.1, 1, 5 for $\beta =$ 3, 5, 7, from upper to lower curves. The crosses are the numerical solutions, the solid lines are the analytical solutions given by Eq. (\ref{eq:barenblatt_solnorm}).}
   \label{fig:barenblatt}
\end{figure*}

\subsubsection{Barenblatt test}

To test our diffusion scheme in the non-linear regime, we performed various computations of the propagation of nonlinear conduction wave, also known as the Barenblatt problem. We solve the equation

\begin{equation}
\label{eq:barenblatt}
\frac{\partial T}{\partial t} =  \alpha \frac{\partial }{\partial x} \Big ( T^\beta \frac{\partial T}{\partial x} \Big ),
\end{equation}

\noindent which is a diffusion equation with a non-linear diffusivity law $\chi = \alpha T^\beta$.

As in the linear diffusion problem, we assume that an instantaneous pulse of energy is released at $x=0$. Let $Q$ be the amplitude of the temperature pulse released in the medium. Following \cite{1984oup..book.....M}, the analytical solution of the problem is

\begin{equation}
\label{eq:barenblatt_sol}
T(x,t) = \Big ( \frac{Q^2}{\alpha t} \Big )^{1/\beta+2} f \Big( \frac{x}{\alpha Q^\beta t} \Big),
\end{equation}

\noindent with
\begin{align}
 f(\xi) &= \Big ( \frac{\beta \xi_0^2}{2(\beta+2)}\Big )^{1/\beta} \Big [ 1 - \Big( \frac{\xi}{\xi_0}\Big )^2 \Big ]^{1/\beta} \mathrm{\ \ \ for\ } \xi < \xi_0\\
 &= 0 \mathrm{,\ otherwise.}
 \end{align}
 
 \noindent $\xi_0$ is defined by
 
 \begin{equation}
 \xi_0 = \Big [ \frac{(\beta+2)^{\beta+1} 2^{1-\beta}}{\beta \pi^{\beta/2}} \frac{\Gamma(\frac{1}{2} + 1/\beta)}{\Gamma(1/\beta)} \Big ]^{1/\beta+2}.
 \end{equation}
  
\noindent The problem can be cast in dimensionless variables, defined in the following way:

\begin{align}
\bar{t} &= \alpha Q^\beta t \\
\bar{x} &= x / \xi_0 \\
\bar{T} &= T / Q f(0).
\end{align}

\noindent In these variables, Eq. (\ref{eq:barenblatt}) becomes

\begin{equation}
\label{eq:barenblattnorm}
\frac{\partial \bar{T}}{\partial \bar{t}} =  \frac{\beta}{2(\beta+2)} \frac{\partial }{\partial \bar{x}} \Big ( \bar{T}^\beta \frac{\partial \bar{T}}{\partial \bar{x}} \Big ),
\end{equation}

\noindent and Eq. (\ref{eq:barenblatt_sol}) now reads as

\begin{equation}
\label{eq:barenblatt_solnorm}
\bar{T}(\bar{x},\bar{t}) = \bar{t}^{-1/\beta+2} \Big [ 1 - \Big( \frac{\bar{x}}{\bar{t}^{1/\beta+2}}\Big)^2 \Big ]^{1/\beta}.
\end{equation}

The advantage is that, in the dimensionless form, the only parameter left is $\beta$. We performed runs for $\beta=$ 1, 3, 5, 7, going from mildly to strongly non-linear regimes. The problem is solved in the dimensionless form for the domain [-1.5,1.5]. Our spatial discretisation uses 100 grid points, and again the time step is set so that the CFL number is equal to 1. As we cannot represent a $\delta$ pulse on a discrete grid accurately, the initial condition in each case is taken to be the analytical solution (\ref{eq:barenblatt_solnorm}) at $\bar{t} = 0.1$.  Figure \ref{fig:barenblatt} shows the results at different times and demonstrates the ability of our code to resolve and track the heat front even for strongly non-linear cases.

\section{Formulae}
\label{appendix:formulae}

\begin{table*}[p]

\begin{align}
\label{CompleteODE}
%
% Continuity equation
%
\frac{d}{d t} \rho_{i+1/2,j+1/2} V_{i+1/2,j+1/2} = - \big( &S^r_{i+1,j+1/2} \tilde{\rho}_{i+1,j+1/2} u^r_{i+1,j+1/2} - S^r_{i,j+1/2} \tilde{\rho}_{i,j+1/2} u^r_{i,j+1/2}\\
+ &S^\theta_{i+1/2,j+1} \tilde{\rho}_{i+1/2,j+1} u^\theta_{i+1/2,j+1} - S^\theta_{i} \tilde{\rho}_{i+1/2,j} u^\theta_{i+1/2,j}\big ) \nonumber \\
%
% Internal energy equation
%
 \frac{d}{d t} \rho_{i+1/2,j+1/2} e_{i+1/2,j+1/2}V_{i+1/2,j+1/2} = - \big( &S^r_{i+1,j+1/2} \widetilde{\rho e}_{i+1,j+1/2} u^r_{i+1,j+1/2} - S^r_{i,j+1/2} \widetilde{\rho e}_{i,j+1/2} u^r_{i,j+1/2} \\
 + &S^\theta_{i+1/2,j+1} \widetilde{\rho e}_{i+1/2,j+1} u^\theta_{i+1/2,j+1} - S^\theta_{i+1/2,j} \widetilde{\rho e}_{i+1/2,j} u^\theta_{i+1/2,j} \big ) \nonumber \\
- P_{i+1/2,j+1/2} (S^r_{i+1,j+1/2} u^r_{i+1,j+1/2} - S^r_{i,j+1/2} u^r_{i,j+1/2} + &S^\theta_{i+1/2,j+1} u^\theta_{i+1/2,j+1} - S^\theta_{i+1/2,j} u^\theta_{i+1/2,j}) \nonumber \\
+ \frac{ac}{3} \Big( S^r_{i+1,j+1/2} \frac{1}{\overline{\rho \kappa}_{i+1,j+1/2}}\frac{T_{i+3/2,j+1/2}^4-T_{i+1/2,j+1/2}^4}{\Delta_{i+1} r} - &S^r_{i,j+1/2} \frac{1}{\overline{\rho \kappa}_{i,j+1/2}} \frac{T_{i+1/2,j+1/2}^4-T_{i-1/2,j+1/2}^4}{\Delta_{i} r} \nonumber \\
+ S^\theta_{i+1/2,j+1} \frac{1}{\overline{\rho \kappa}_{i+1/2,j+1}}  \frac{T_{i+1/2,j+3/2}^4-T_{i+1/2,j+1/2}^4}{r_{i+1/2} \Delta_{j+1} \theta} - &S^\theta_{i+1/2,j} \frac{1}{\overline{\rho \kappa}_{i+1/2,j}} \frac{T_{i+1/2,j+1/2}^4-T_{i+1/2,j-1/2}^4}{r_{i+1/2} \Delta_{j} \theta} \Big ) \nonumber \\
%
% Radial momentum equation
%
\frac{d}{d t} \bar{\rho}_{i,j+1/2} u^r_{i,j+1/2} \hat{V}_{i,j+1/2} = - \big( &\hat{S}^r_{i+1/2,j+1/2} \widetilde{\rho u^r}_{i+1/2,j+1/2} \bar{u}^r_{i+1/2,j+1/2}\\
 &- \hat{S}^r_{i-1/2,j+1/2}\widetilde{\rho u^r}_{i-1/2,j+1/2} \bar{u}^r_{i-1/2,j+1/2} \nonumber \\
& + \hat{S}^\theta_{i,j+1} \widetilde{\rho u^r}_{i,j+1} \bar{u}^\theta_{i,j+1} - \hat{S}^\theta_{i,j}\widetilde{\rho u^r}_{i,j} \bar{u}^\theta_{i,j} \big ) \nonumber \\
 - &\frac{P_{i+1/2,j+1/2} - P_{i-1/2,j+1/2}}{\Delta_i r}\hat{V}_{i,j+1/2} + \bar{\rho}_{i,j+1/2} g_{i,j+1/2} \hat{V}_{i,j+1/2} \nonumber \\
%+ &\hat{V}_{i,j+1/2} \frac{ \overline{\rho u}^\theta_{i,j+1/2} \bar{u}^\theta_{i,j+1/2} }{r_i} \nonumber \\
+ &\hat{V}_{i,j+1/2} \frac{ \bar{\rho}_{i,j+1/2} (\bar{u}^\theta_{i,j+1/2})^2 }{r_i} \nonumber \\
%
% Tangential momentum equation
%
\frac{d}{d t} \bar{\rho}_{i+1/2,j} u^\theta_{i+1/2,j} \check{V}_{i+1/2,j} = - \big( &\check{S}^r_{i+1,j} \widetilde{\rho u^\theta}_{i+1,j} \bar{u}^r_{i+1,j} - \check{S}^r_{i,j}\widetilde{\rho u^\theta}_{i,j} \bar{u}^r_{i,j} \\
& + \check{S}^\theta_{i+1/2,j+1/2} \widetilde{\rho u^\theta}_{i+1/2,j+1/2} \bar{u}^\theta_{i+1/2,j+1/2} \nonumber \\
&  - \check{S}^\theta_{i+1/2,j-1/2}\widetilde{\rho u^\theta}_{i+1/2,j-1/2} \bar{u}^\theta_{i+1/2,j-1/2} \big ) \nonumber \\
 - &\frac{P_{i+1/2,j+1/2} - P_{i+1/2,j-1/2}}{r_{i+1/2}\Delta_j \theta}\check{V}_{i+1/2,j} 
% - \check{V}_{i+1/2,j} \frac{ \overline{\rho u}^r_{i+1/2,j}  u^\theta_{i+1/2,j} }{r_{i+1/2}} \nonumber \\
 - \check{V}_{i+1/2,j} \frac{ \bar{\rho}_{i+1/2,j}  \bar{u}^r_{i+1/2,j}  u^\theta_{i+1/2,j} }{r_{i+1/2}}  \nonumber \\
\end{align}
 
 \begin{align}
 % Mean conductivity
\frac{1}{\overline{\rho \kappa}_{i,j+1/2}} &=  \frac{1}{2} (\frac{1}{\rho_{i-1/2,j+1/2} \kappa_{i-1/2,j+1/2}} + \frac{1}{\rho_{i+1/2,j+1/2} \kappa_{i+1/2,j+1/2}}) &  & \\
\frac{1}{\overline{\rho \kappa}_{i+1/2,j}} &=  \frac{1}{2} (\frac{1}{\rho_{i+1/2,j-1/2} \kappa_{i+1/2,j-1/2}} + \frac{1}{\rho_{i-1/2,j+1/2} \kappa_{i+1/2,j+1/2}}) &  & \\
 % Mean density
 \bar{\rho}_{i,j+1/2} &= (\rho_{i-1/2,j+1/2} V_{i-1/2,j+1/2} + \rho_{i+1/2,j+1/2} V_{i+1/2,j+1/2})/\hat{V}_{i,j+1/2} & & \\
 \bar{\rho}_{i+1/2,j} &= (\rho_{i+1/2,j-1/2} V_{i+1/2,j+1/2} + \rho_{i+1/2,j-1/2} V_{i+1/2,j-1/2})/\check{V}_{i+1/2,j} & & \\
 % Mean velocity
 \bar{u}^r_{i+1/2,j+1/2} &= \frac{1}{2}(u^r_{i,j+1/2} + u^r_{i+1,j+1/2}) & \bar{u}^\theta_{i,j} &= \frac{1}{2}(u^\theta_{i-1/2,j} + u^\theta_{i+1/2,j} )\\
 \bar{u}^r_{i,j} &= \frac{1}{2}(u^r_{i,j-1/2} + u^r_{i,j+1/2}) & \bar{u}^\theta_{i+1/2,j+1/2} &= \frac{1}{2}(u^\theta_{i+1/2,j} + u^\theta_{i+1/2,j+1})\\
  \bar{u}^r_{i+1/2,j} &= \frac{1}{2}(\bar{u}^r_{i,j} + \bar{u}^r_{i+1,j}) & \bar{u}^\theta_{i,j+1/2} &= \frac{1}{2}(\bar{u}^\theta_{i,j} + \bar{u}^\theta_{i,j+1})\\
 \end{align}
 
\caption{Complete system of ordinary differential equations resulting from our finite volume discretisation on a staggered grid in spherical coordinates.}
\end{table*}

We recall the geometrical factors for the scalar, cell centred, quantities:

\begin{align}
V_{i+1/2,j+1/2} &= \frac{2\pi}{3}(r_{i+1}^3 - r_i^3)(\cos \theta_j - \cos \theta_{j+1}) \nonumber \\
S^r_{i,j+1/2} &= 2 \pi r_i^2(\cos \theta_j - \cos \theta_{j+1}) \nonumber \\
S^\theta_{i+1/2,j} &= \pi \sin \theta_j (r_{i+1}^2-r_{i}^2).
\end{align}

\noindent Geometrical factors for the radial velocity
\begin{align}
\hat{V}_{i,j+1/2} &= \frac{2\pi}{3}(r_{i+1/2}^3 - r_{i-1/2}^3)(\cos \theta_j - \cos \theta_{j+1}) \nonumber \\
\hat{S}^r_{i-1/2,j+1/2} &= 2 \pi r_{i-1/2}^2(\cos \theta_j - \cos \theta_{j+1}) \nonumber \\
\hat{S}^\theta_{i,j} &= \pi \sin \theta_j (r_{i+1/2}^2-r_{i-1/2}^2).
\end{align}

\noindent Geometrical factors for the tangential velocity
\begin{align}
\check{V}_{i+1/2,j} &= \frac{2\pi}{3}(r_{i+1}^3 - r_i^3)(\cos \theta_{j-1/2} - \cos \theta_{j+1/2}) \nonumber \\
\check{S}^r_{i,j} &= 2 \pi r_i^2(\cos \theta_{j-1/2} - \cos \theta_{j+1/2}) \nonumber \\
\check{S}^\theta_{i+1/2,j-1/2} &= \pi \sin \theta_{j-1/2} (r_{i+1}^2-r_{i}^2).
\end{align}

\noindent The complete set of ordinary differential equations resulting from our spatial discretisation in spherical coordinates are summarised in Tab. \ref{CompleteODE}.

\end{document}